%% file: main.tex
\documentclass[twocolumn]{aastex63}

\usepackage{graphicx}
\usepackage{longtable}
\usepackage{url}
\usepackage{hyperref}
\hypersetup{colorlinks=false,pdfborder={0 0 0}}

\usepackage[utf8]{inputenc}
\usepackage[english]{babel}
\newcommand{\minus}{\scalebox{0.3}[1.0]{$-$}}

\usepackage{color}
\usepackage[normalem]{ulem}

\shorttitle{APOGEE/\emph{Kepler} Overlap}
\shortauthors{J. M. C. Cunningham et al.}

\begin{document}

\title{APOGEE/\emph{Kepler} Overlap Yields Orbital Solutions for a Variety of Eclipsing Binaries}

\author{Joni Marie Clark Cunningham}
\affil{Department of Physics, Fisk University, Nashville, TN 37208}
\affil{Department of Physics and Astronomy, Vanderbilt University, Nashville, TN 37235}

\author{Meredith L. Rawls}
\affiliation{Department of Astronomy, University of Washington, Seattle, WA 98195}
\affiliation{DIRAC Institute, University of Washington, Seattle, WA 98195}

\author{Diana Windemuth}
\affiliation{Department of Astronomy, University of Washington, Seattle, WA 98195}

\author{Aleezah Ali}
\affiliation{Department of Astronomy, University of Washington, Seattle, WA 98195}

\author{Jason Jackiewicz}
\affiliation{Department of Astronomy, New Mexico State University, Las Cruces, NM 88003}

\author{Eric Agol}
\affiliation{Department of Astronomy, University of Washington, Seattle, WA 98195}

\author{Keivan G. Stassun}
\affiliation{Department of Physics and Astronomy, Vanderbilt University, Nashville, TN 37235}
\affil{Department of Physics, Fisk University, Nashville, TN 37208}

\begin{abstract}

Spectroscopic Eclipsing Binaries (SEBs) are fundamental benchmarks in stellar astrophysics and today are observed in breathtaking detail by missions like TESS, \emph{Kepler}, and APOGEE. We develop a methodology for simultaneous analysis of high precision \emph{Kepler} light curves and high resolution near-IR spectra from APOGEE and present orbital solutions and evolutionary histories for a subset of SEBs within this overlap. Radial velocities extracted from APOGEE spectra using the Broadening Function technique are combined with \emph{Kepler} light curves and to yield binary orbital solutions. The Broadening Function approach yields more precise radial velocities than the standard Cross-Correlation Function, which in turn yields more precise orbital parameters and enables the identification of tertiary stars. The orbital periods of these seven SEBs range from 4 to 40 days. Four of the systems (KIC 5285607, KIC 6864859, KIC 6778289, and KIC 4285087) are well-detached binaries. The remaining three systems have apparent tertiary companions, but each exhibits two eclipses along with at least one spectroscopically varying component (KIC 6449358, KIC 6131659, and KIC 6781535). \emph{Gaia} distances are available for four targets which we use to estimate temperatures of both members of these SEBs. We explore evolutionary histories in H-R diagram space and estimate ages for this subset of our sample. Finally, we consider the implications for the formation pathways of close binary systems via interactions with tertiary companions. 
Our methodology combined with the era of big data and observation overlap opens up the possibility of discovering and analyzing large numbers of diverse SEBs, including those with high flux ratios and those in triple systems. 
\end{abstract}


\section{Introduction}
\label{intro}

The Apache Point Observatory Galactic Evolution Experiment (APOGEE) is studying our Galaxy in fantastic detail by providing high resolution spectra for some 150,000 stars \citep{Majewski_2015}. Some of these belong to double-lined spectroscopic eclipsing binaries (SEBs), and a further subset have been observed by the \emph{Kepler} spacecraft \citep{Borucki_2010} and appear in the \emph{Kepler} Eclipsing Binary Catalog \citep{Kirk_2016}. These APOGEE/\emph{Kepler} SEBs which have several APOGEE spectra at different epochs give a unique opportunity to combine the spectra with the \emph{Kepler} light curve to model the binary orbit and directly measure fundamental stellar parameters, including mass and radius. They can then be used to explore and constrain stellar evolution, stellar populations, and orbital kinematics.   

While much work has gone into exploring \emph{Kepler} eclipsing binaries (EBs) as a population, fewer studies have maximally utilized complementary spectra to fully characterize these stellar systems. A notable exception is \citet{Matson_2017}, which found the radial velocities of 40 \emph{Kepler} binaries, 35 of them double-lined and the remainder single-lined. Their work used medium resolution ground-based spectra, but the authors note that high resolution spectra is more optimal. In another example, \citet{Torres_2018} used \emph{K2} light curves of the Ruprecht 147 cluster together with high resolution spectra and the cluster's well-modeled metallicity to constrain the orbital parameters extracted from spectroscopic binary cluster members. In addition, \citet{Lehmann_2012} analyzed the quadruple system KIC 4247791 by combining \emph{Kepler} light curves and moderate resolution spectra.

Many studies have used APOGEE and \emph{Kepler} data together, such as the APOKASC catalog \citep{Pinsonneault_2018, Pinsonneault_2014} which combines APOGEE stellar parameters with \emph{Kepler} asteroseismology. However, such works tend to ignore stellar multiplicity. The SEB overlap between APOGEE and \emph{Kepler} in particular remains relatively unexplored.  There is also synergy with the \emph{Kepler} planet survey which identifies candidate planet systems, some of which are found to be eclipsing binaries, with or without tertiaries, with follow-up radial-velocity observations with APOGEE \citep{Fleming2015}.  The frequency of binaries, with and without tertiary companions, is a necessary component of computing transiting exoplanet astrophysical false-alarm probabilities \citep[e.g.][]{Morton2016}. 

EBs have long been used as fundamental benchmarks for stellar astrophysics \citep[e.g.,][]{Torres:2010}, including more recently as benchmarks for exoplanet properties \citep[e.g.,][]{Stassun:2017}, to test asteroseismic inferences of stellar parameters \citep[e.g.,][]{Gaulme:2016}, and even for assessing trigonometric parallaxes \citep{Stassun:2018}. In addition, as they are often observed as SEBs, EBs are useful for assembling reliable statistics on the occurrence of higher order multiples (e.g., tertiary companions) and on the relationship of companion properties to the properties of the EB. For example, \citet{Tokovinin:1997} found the incidence of wide tertiaries to be strongly linked to the orbital period of the inner binary. Additional well-studied EBs can help to further test these relationships.

In this work, we identify 33 promising APOGEE/\emph{Kepler} SEBs and compute full orbital solutions with a suite of stellar parameters for seven of them. In \S \ref{data}, we detail our sample selection, data processing, and modeling methodology. We further show how the Broadening Function technique is a superior method to extract multiple velocity components from APOGEE spectra. Subsequently \S \ref{results} discusses each of the seven modeled systems in turn and presents orbital solutions. Finally, \S \ref{discuss} places the SEBs in the context of each star's stellar evolutionary history and explores the relationship of the EB orbital properties to the presence of tertiary companions.

\section{Data and Methods}
\label{data}

\subsection{Sample Selection}
\label{sample}

We use the following criteria and filters to arrive at a candidate sample of promising SEBs in the APOGEE/\emph{Kepler} overlap. We begin with the \emph{Kepler} EB catalog compiled by \citet{Kirk_2016}. From this catalog we select targets which have both their primary and secondary eclipses observed by \emph{Kepler}; this limits our selection to binaries with inclinations close to 90 degrees. We further require the light curve to be semi- or well-detached, with the morphology parameter significantly less than 1. Next, a luminosity limit of $H < 14$ magnitudes was imposed, as fainter targets are unlikely to have $H$-band APOGEE spectra with a sufficiently high signal-to-noise ratio. We also require the targets to have multiple cross-correlation function (CCF) peaks from the APOGEE pipeline \citep{Nidever_2015} visible by eye in one epoch. Finally, the binaries must have been observed by APOGEE at least three times, and thus have at least three \texttt{apVisit} spectra, with no quality flags present.

Taken together, these criteria result in 33 candidates, which are listed in Table \ref{table1}, plus one additional candidate which has already been analyzed \citep{Rawls_2016}. Of these, we perform a detailed analysis of seven. Notes in Table \ref{table1} indicate why we choose to exclude the other systems at this time. Several are being investigated by the \emph{Kepler} APOGEE EB Working Group, some have only three APOGEE visits which would make a good RV curve solution challenging without additional spectra, some have low signal-to-noise (S/N) ratios, one shows significant ellipsoidal variations which are not included in our photometric model, and two remain good candidates for future analyses.

\input{table1.tex}  

\subsection{Radial Velocities (RVs) from APOGEE
Spectra}
\label{apogee}

The standard observing mode for APOGEE spectra has a total exposure time of roughly three hours, which is usually collected over a series of visits on different days. The visits are then combined into one spectrum per target (an \texttt{apStar} spectrum). We instead utilize individual visit spectra (\texttt{apVisit}), which are identified with their plate ID, date (MJD), and fiber ID. These may be retrieved from the SDSS Science Archive Server search tool with a simple search by APOGEE ID.
We continuum normalize the visit spectra and then ``de-spike" them to remove erroneous spectral features due to tellurics. De-spiking consists of identifying outliers above or below the continuum by 0.7 or 3 times the standard deviation of the normalized flux, respectively. The ``below continuum'' factor is larger to avoid unintentionally removing real absorption line features. Around each outlier spike, a $\pm 6$ \AA \ window is also flagged for removal. The python scripts used to retrieve, continuum normalize, and de-spike \texttt{apVisit} spectra are publicly available~\href{https://github.com/mrawls/apVisitproc}{on GitHub}\footnote{\texttt{https://github.com/mrawls/apVisitproc}}. They rely heavily on the~\href{https://github.com/jobovy/apogee}{apogee python package on GitHub}~described in \citet{Bovy_2016}.

In the main APOGEE reduction pipeline \citep{Nidever_2015}, RVs are
measured using the CCF. In this approach, a
template spectrum and a series of visit spectra for a given target are
cross-correlated, giving the RV of the target star relative to the template.

The CCF method works because an observed stellar spectrum can be represented as a convolution of two functions: that of the astrophysical
target (which includes ``natural broadening'' components such as thermal
broadening, microturbulence effects, and instrumental broadening) and
another function called the Broadening Function (BF) that contains the important RV information. The BF is formally presented
in \citet{Rucinski_1992,Rucinski_1999,Rucinski_2002,Rucinski_2004}.
Although the  CCF method is very close to the real convolution that occurs in an APOGEE spectrum, cross-correlating a template
spectrum with an observed stellar spectrum yields a function which
inherits the natural broadening components present in both spectra. In
this way the CCF is essentially a non-linear proxy of the BF. Therefore, instead of using the CCF, in this work we measure the BF directly.

To extract BFs from our target spectra, we use a modified version of the BF software suite from \citet{Rawls_2016} which is based on the method introduced by \citet{Rucinski_1992}. A PHOENIX BT-Settl model atmosphere spectrum \citep{Husser_2013} is selected to match the target's approximate spectral parameters as reported by APOGEE. The match cannot be exact as the two stars in a binary may not have identical spectral types and the model grid has a finite sampling in stellar parameters. In general, a mismatch in spectral type between template and target causes the BF to change in its intensity scale and quality, but the amplitudes of the RV components remain unchanged \citep{Lu_2001}.

We examine the BF peaks by eye to identify their approximate locations on the radial velocity axis and use a least-squares fitting procedure to fit one or more Gaussians to the BF. The location of each Gaussian's mean is the RV, which we then correct with the barycentric velocity provided with each \texttt{apVisit} spectrum. Our reported RV uncertainties come from the error in fitting a Gaussian to each BF peak using least-squares. Much like in the APOGEE CCF pipeline \citep{Nidever_2015}, it is ultimately the uncertainty in the measurement of the BF peak, which depends partially on its semi-arbitrary width, that determines the reported uncertainties for the RVs. This systematically underestimates the uncertainty of each RV measurement. The software used to extract RVs as described here is publicly available \href{https://github.com/savvytruffle/cauldron/tree/master/rvs}{on GitHub}.

\begin{figure*}[ht!]
\begin{center}
\includegraphics[width=1.80\columnwidth]{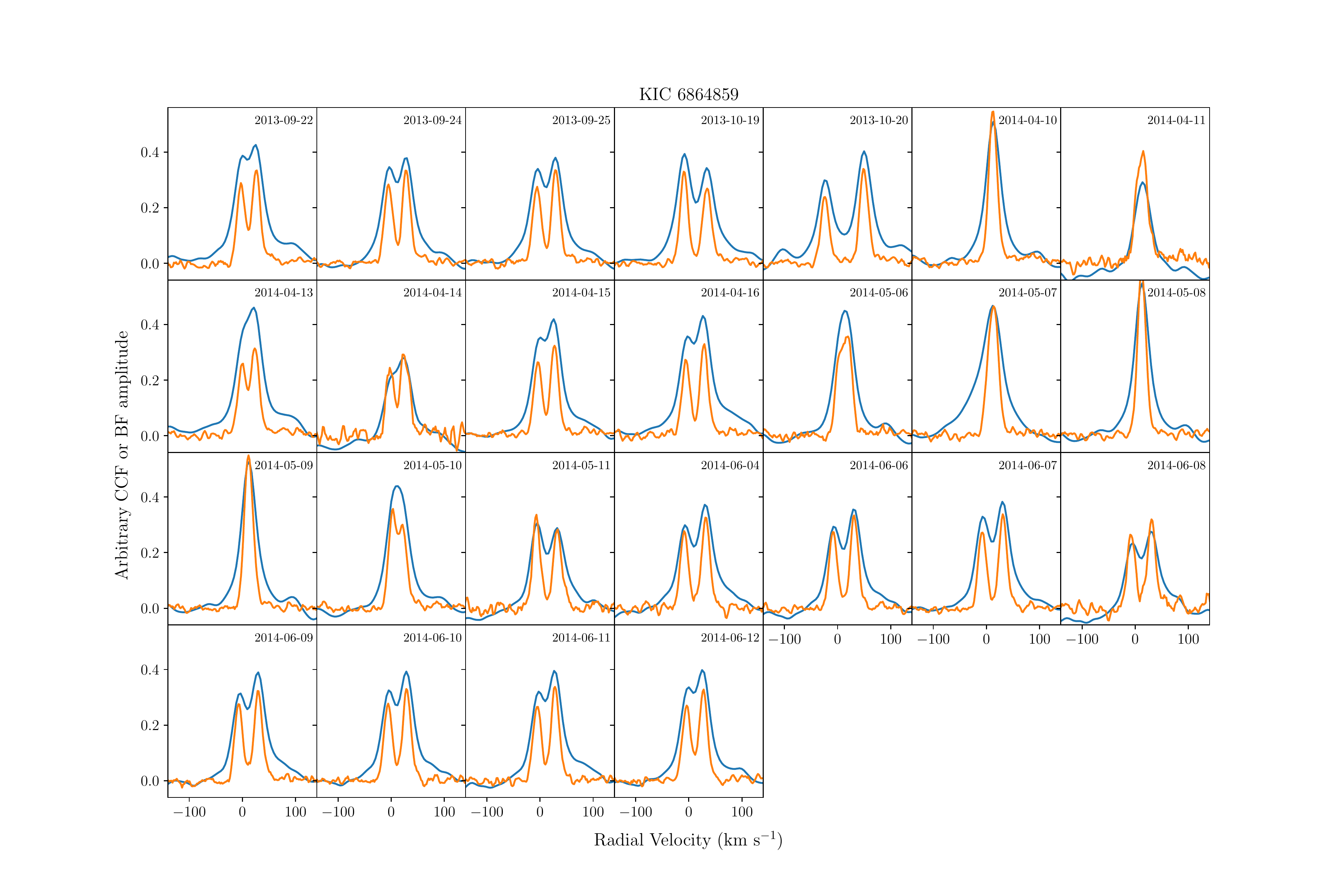}
\caption{{The BF (orange) is significantly better at resolving multiple velocity components from APOGEE visit spectra than the CCF (blue). This example shows visits for KIC 6864859. The y-axis amplitude is arbitrarily scaled for clarity. While it is clear from most of the CCFs that KIC 6864859 is a double-lined SEB, the BF more clearly separates the contribution from each star.
{\label{fig_bfcompare1}}
}}
\end{center}
\end{figure*}

\begin{figure*}[ht!]
\begin{center}
\includegraphics[width=1.80\columnwidth]{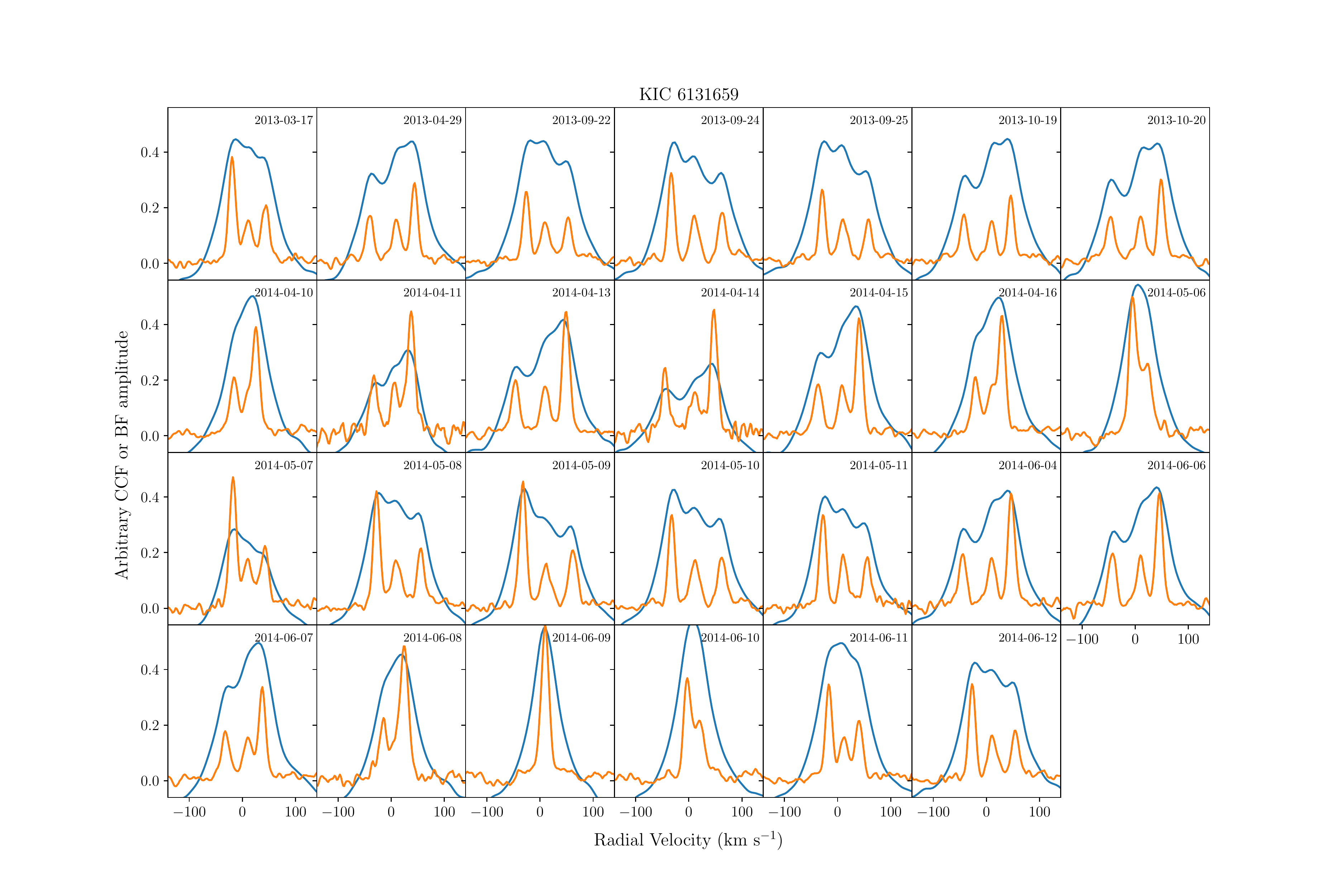}
\caption{{A second example, as in Figure \ref{fig_bfcompare1}, showing that the BF (orange) does a better job of resolving multiple velocity components than the CCF (blue). The y-axis amplitude is arbitrarily scaled for clarity. While it is clear from some of the CCFs that KIC 6131659 has more than one component, the three-component nature is immediately obvious in the BF. In all but a few cases, it is straightforward to precisely measure the RV of all three BF peaks. This is not true for any visit using the APOGEE CCFs.
{\label{fig_bfcompare2}}
}}
\end{center}
\end{figure*}

In Figures \ref{fig_bfcompare1} and \ref{fig_bfcompare2}, we demonstrate how the BF method produces significantly better separated peaks for APOGEE double-lined SEBs than the CCFs generated by the APOGEE pipeline. Due in part to the BF method having less of a ``peak-pulling" effect, this more defined separation dramatically improves our ability to  measure the RV of each component. We present the BF and measured RVs for each of our seven targets in Appendix \ref{appendixA}.

We also measure flux ratios in the APOGEE $H$ band with our BF peaks, as the ratio of the peak areas is directly proportional to the flux ratio of the binary \citep{Bayless_2006,Stassun_2007}. These BF flux ratios will generally differ from \emph{Kepler}-derived flux ratios because of the difference in wavelength; APOGEE is an $H$-band spectrograph and \emph{Kepler} has a broad visible light bandpass. The BF flux ratios are discussed further in Section \ref{supp_parms} alongside other variables for RV extraction and temperature estimation.

\subsection{Kepler Light Curve Processing}
\label{kepler}

We use minimally-processed \emph{Kepler} long-cadence simple aperture photometry (SAP) to construct EB light curves for each target. Each light curve, and its uncertainties, is normalized by the median raw flux value of all available quarters. We de-weight data of poor quality by selecting times with \emph{Kepler} quality flags \textgreater{} 0, and inflating the normalized flux uncertainties at these times by a factor of 10.

\subsection{Orbital and Mass Solutions with KEBLAT}
\label{keblat}

With RVs in hand, we turn to the photometric modeling of the \emph{Kepler} light curves.  We utilize a modular Python tool dubbed ``KEBLAT" which is capable of separate or simultaneous modeling of the binary light curve, spectral energy distribution (SED), and RV time series \citep{Windemuth_2018}. Here, we simultaneously model \emph{Kepler} light curves and APOGEE RVs of each EB in our sample to determine orbital solutions
($P,\ e,\ \omega,\ i,\ t_{PE}$),
stellar parameters
($m_1,\ m_2,\ r_1,\ r_2,\ \frac{F_2}{F_1}$),
quadratic limb darkening coefficients under triangular reparameterization ($q_{1,1},\ q_{1,2},\ q_{2,1},\ q_{2,2}$; \citealt{Kipping_2013}),
and systemic radial velocity $k_0$. For parameter sampling purposes, we transform individual mass and radius parameters to sums and ratios, and parameterize $e$ and $\omega$ as $e\cos \omega$ and $e\sin \omega$.

Given a system's total mass, period, eccentricity, argument of periastron, inclination, and time of primary eclipse, KEBLAT uses a Keplerian solver to compute the instantaneous positions and velocities of each stellar component. The positions, along with specified sizes and relative flux of the stars are then used to determine the instantaneous light contribution during eclipse via a quadratic limb-darkening \citep{Mandel_2002} model for spherical bodies.\footnote{The assumption of spherical stars requires that the stars be sufficiently detached to avoid tidal and rotational distortions.} We account for finite sampling effects \citep{Kipping_2010} on the light curve by down-sampling 1-minute eclipse profiles to the \emph{Kepler} long cadence $\left(dt=0.0204\ \textrm{d} \right)$. Stellar and instrumental noise is marginalized by fitting the lowest non-linear order quadratic polynomial around each eclipse. We apply quarterly crowding values from \emph{Kepler} to model third light contamination. To account for underestimated observational uncertainties and additional noise, we fit for a systematic light curve error $\sigma_{\mathrm{sys, LC}}$, which we add in quadrature to the observed errors. 

The z-component of the velocity, as solved by Kepler's equation, is used to model the extracted RVs. For double-line eclipsing binary systems, where the RVs of both components are detected, the amplitudes of the primary and secondary RV are related to the masses of the secondary and primary, respectively. For single-line EBs, where only the RV of the brightest component is detected, only the ``mass function" $f_M$ of the system can be constrained, where
\begin{equation}
f_M = \frac{M_2^3 \sin^3 i}{(M_1+M_2)^2}.
\end{equation}
As with the light curve data, we fit for a systematic radial velocity error parameter $\sigma_{\mathrm{sys, RV}}$ to account for underestimated noise. 

We combine RVs with \emph{Kepler} light curve information to model the system and find a best fit solution. We first determine the light curve and RV solutions separately, and then fit RV and light curve simultaneously.  The simultaneous RV+LC model has 17 free parameters in total. The model is optimized via a least-squares algorithm \texttt{lmfit} \citep{Newville_2014}, and then uses the best-fit solution to seed Monte Carlo Markov Chain (MCMC) simulations with \texttt{emcee} \citep{Foreman-Mackey_2013}, in order to sample the posterior distributions of each parameter. We use broad, uniform priors and run the Markov chains with 128 walkers for $\sim$100,000 iterations, visually inspecting trace plots for convergence. We report the 50\%, 16\%, and 86\% values, i.e., the mean and $1 \sigma$ uncertainties for each parameter. For more details on the KEBLAT model, including parameter bounds, see \citet{Windemuth_2018}.

\subsection{Radius ratio---flux ratio---inclination Degeneracy}
\label{degeneracy}

For light curves with partially or grazing eclipsing geometries, there exists a degeneracy between radius ratio and flux ratio when eclipses are observed in a single photometric band. For this reason, we use additional constraints on the \emph{Kepler} light curve flux ratios with spectroscopic $H$-band flux ratios obtained from the BF for SEBs exhibiting shallow eclipses (KIC 5285607 and KIC 6781535). For these two systems, we place a Gaussian prior on the RV+LC solution with $\mu$ centered around the BF-derived flux ratio and $\sigma=0.2$. 

\subsection{Temperatures from Flux Ratios and Radii}
\label{temps}

Obtaining effective temperatures of the stars in these SEBs requires additional analysis. The KEBLAT model does not provide a measure of stellar temperatures directly, but only indirectly via the flux ratio in the \emph{Kepler} bandpass. In addition, the APOGEE Stellar Parameter and Chemical Abundances Pipeline (ASPCAP) processing reports only a single ``combined light" effective temperature for each system \citep{Garcia-Perez_2016}. This ASPCAP temperature is likely to be biased due to the contamination of the brighter star's spectrum by the fainter star. Here we address both of these issues to estimate individual stellar effective temperatures.

The orbital solutions described in \S \ref{keblat} yield sums and ratios of radii.
In addition, the light curve analyses yield flux ratios ($F_{\textrm{ratio}} \equiv \frac{F_2}{F_1}$) in the \emph{Kepler} bandpass which are primarily constrained by the observed eclipse depths.

By assuming the ASPCAP effective temperature is the flux weighted average of the system, $T_{\textrm{avg}} = (F_1 T_1 + F_2 T_2)/(F_1 + F_2)$, and defining the primary star as the one that provides the majority of the light, we can use the following relationships between the flux ratio $F_{\textrm{ratio}}$, $T_{\textrm{avg}}$, stellar radii $R_1$ and $R_2$, and distance $d$ to find flux and temperature estimates for the individual binary components separately. First, we solve for the binary's flux sum using \emph{Gaia} distance estimates from \citet{Bailer-Jones_2018}:
\begin{equation}
F_{\textrm{sum}} = \frac{\sigma T_{\textrm{avg}}^4 (R_{1}^2 + R_{2}^2)}{d^2},
\end{equation}
from which we can compute the individual fluxes as
\begin{equation}
F_{1} = \frac{F_{\textrm{sum}}}{1 + F_{\textrm{ratio}}}
\end{equation}
and
\begin{equation}
F_{2} = F_{\textrm{sum}} - F_{1}.
\end{equation}

Then, we use the relationship between $T_{\textrm{ave}}$, ${F_{\textrm{ratio}}}$, and the individual stellar fluxes to solve for the temperature of each star: 
\begin{equation}
    T_{\textrm{avg}} = T_2 \frac{F_{\textrm{ratio}}^{-1}
    T_{\textrm{ratio}}^{-1}+1}{F_{\textrm{ratio}}^{-1}+1},
\end{equation}
which yields
\begin{equation}
    T_{1} = \sqrt[4]{\frac{F_{1} \ d^2}{\sigma  R_{1}^2}}
\end{equation}
and
\begin{equation}
    T_{2} = \sqrt[4]{\frac{F_{2} \ d^2}{\sigma  R_{2}^2}}.
\end{equation}

However, these resulting temperature estimates are likely systematically underestimated. In the ASPCAP pipeline, APOGEE spectra are compared to a synthetic spectral model to resolve quantities like effective temperature. When a detached binary signature is present in stellar spectra, the additional component can cause the spectrum to be fit by a cooler synthetic template. This leads to a systematic underestimation in the binaries' effective temperatures of roughly 300 K \citep{ElBadry_2017}. The systematic underestimation is a function of the effective temperature of the primary and the mass ratio of the system. We follow their method for each system to correct for this effect.

\section{Results}
\label{results}

The joint light curve and RV analysis for each of the seven systems is presented in detail in the following subsections. We present orbital and mass solutions together in Table \ref{table2}. 

In all systems, we define the primary eclipse ($\phi=0$) as the deeper eclipse. This corresponds to when the primary star is eclipsed by the secondary star. Usually, the primary star is the brighter of the two, and the light curve definition of primary and the RV definition of primary agree. However, we note the secondary star in KIC 6781535 is brighter in the APOGEE $H$-band than the primary. In the figures that follow, we color code the RV of the primary star in red and the RV of the secondary star in orange.

\input{table2.tex}  

\subsection{KIC 5285607}
\label{5285607}

\begin{figure}[ht!]
\begin{center}
\includegraphics[width=1.1\columnwidth]{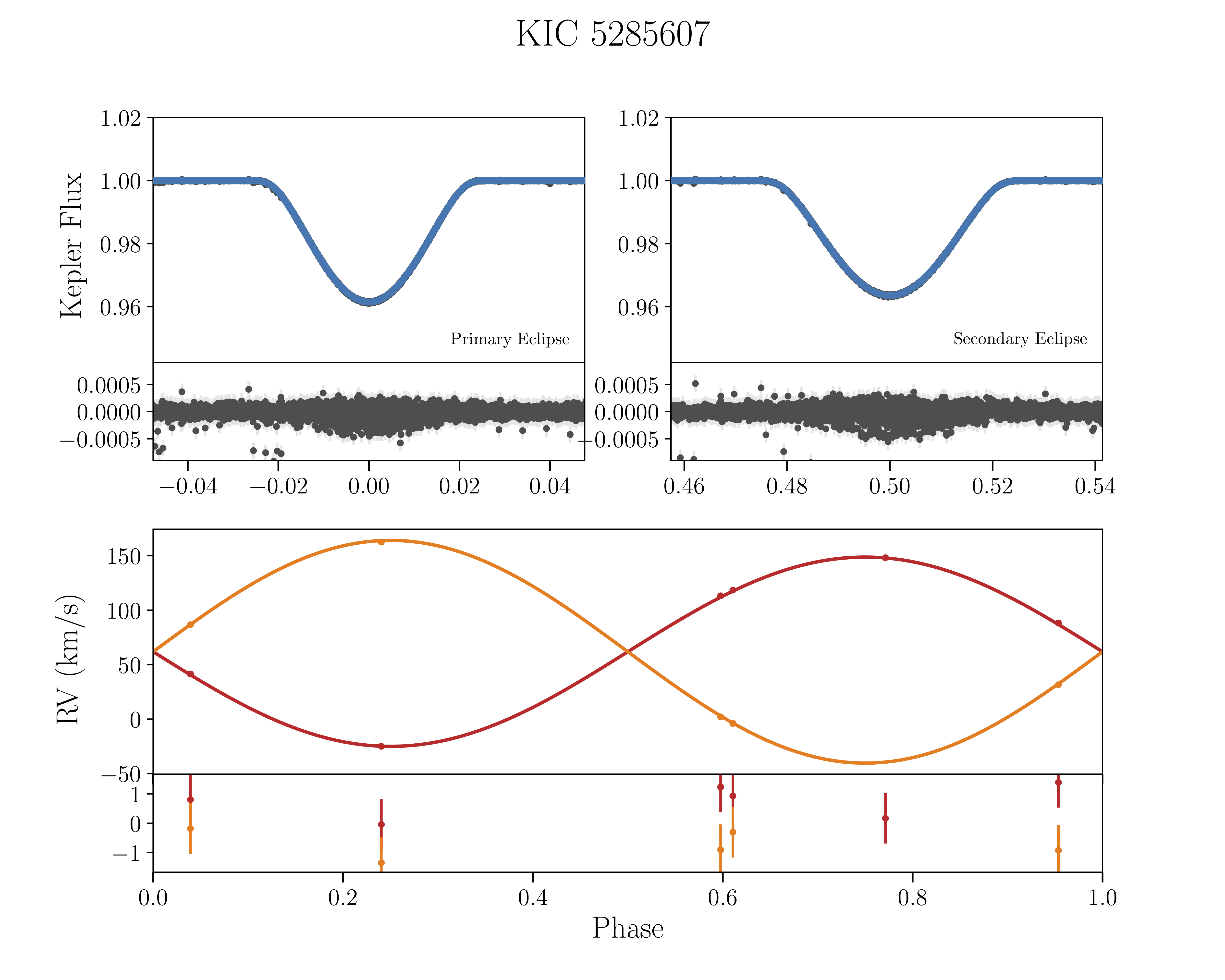}
\caption{{Simultaneous light curve (top panels) and radial velocity (bottom panel) fits to \emph{Kepler} and APOGEE observations for KIC 5285607. The light curve data and residuals are shown in dark grey, with the model overlaid in blue. Note the shallow and V-shaped primary and secondary eclipses, which indicate a grazing system. The phased RV panel shows the BF derived values (points) and corresponding model (lines), where the primary RV is red and secondary RV is orange. The RV semi-amplitude is $\sim$100 km/s while the residual scatter is $\sim$1~km~s$^{-1}$.
{\label{fig_5285607fit}}
}}
\end{center}
\end{figure}

KIC 5285607 is a grazing ($i=79^{\circ}$) 3.9 d eclipsing binary with similar mass stellar
components ($M_1 = 1.56~M_{\odot}, M_2 = 1.35~M_{\odot}$). The stars are in a circular orbit, as exhibited by the sinusoidal shape of the RV and occurrence of secondary eclipse at a phase of $\sim 0.5$ as seen in Figure~\ref{fig_5285607fit}.

Because the eclipses in the light curves are shallow (4\% loss of light), the impact parameter is highly degenerate with the flux and radius ratio. That is, a solution with similar flux contributions from both components in a more inclined system yields the same shallow eclipses as a solution with a much brighter primary component in a more edge-on system. This degeneracy can be ameliorated with additional information from spectra. Therefore, we place a Gaussian prior on the flux ratio parameter based on the BF fits, with a 0.2 1$\sigma$ width. With this flux ratio constraint, we find the secondary star is about 80\% the size of the primary, with absolute dimensions of  $2.0~R_{\odot}$ and $1.7~R_{\odot}$ for the primary and secondary components, respectively. 

\subsection{KIC 6864859}
{\label{6864859}}

\begin{figure}[ht!]
\begin{center}
\includegraphics[width=1.1\columnwidth]{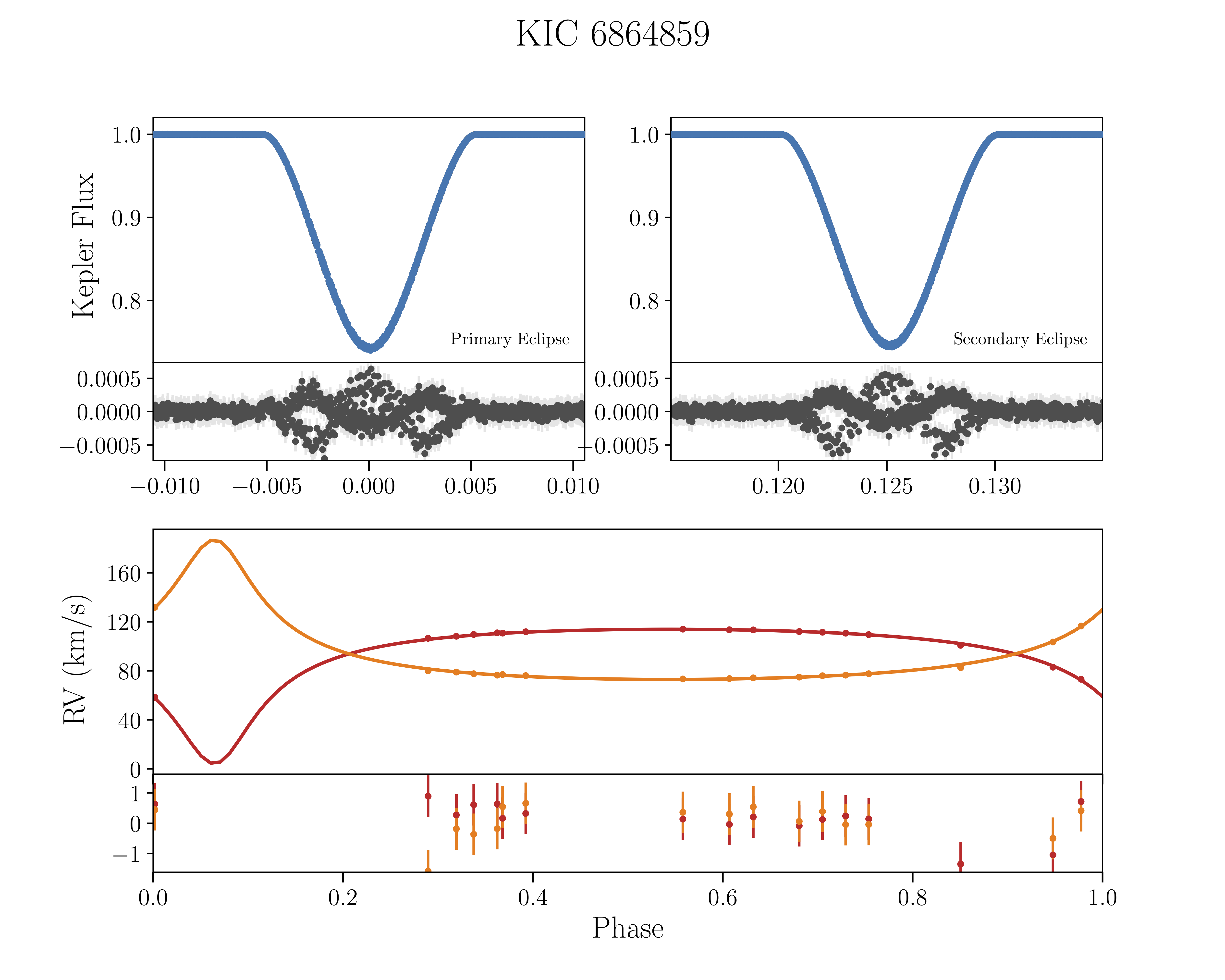}
\caption{{Simultaneous light curve (top panels) and radial velocity (bottom panel) fits to \emph{Kepler} and APOGEE observations of KIC 6864859. The primary and secondary eclipses are similar in shape and depth with $\sim$25\% loss of light; the phase of secondary eclipse and shapes of the RVs indicate an extremely eccentric system. The light curve residuals are small but have a coherent shape, likely due to tidal and rotational distortion of the stars.
{\label{fig_6864859fit}}
}}
\end{center}
\end{figure}

KIC 6864859 is a highly eccentric ($e\sim 0.6$), slightly inclined ($i=88.32^{\circ}$) 40.9 d eclipsing binary with components of similar mass ($M_1=1.41\ M_{\odot}$, $M_2=1.35\ M_{\odot}$) and radii ($R_1 = 1.66 \ R_{\odot}$, $R_2 = 1.46 \ R_{\odot}$). The best-fit model is shown in Figure~\ref{fig_6864859fit}. The system's highly eccentric orbit gives rise to irregularly shaped RVs and a secondary eclipse near phase $\sim$ 0.125.

\begin{figure}[ht!]
\begin{center}
\includegraphics[width=1.1\columnwidth]{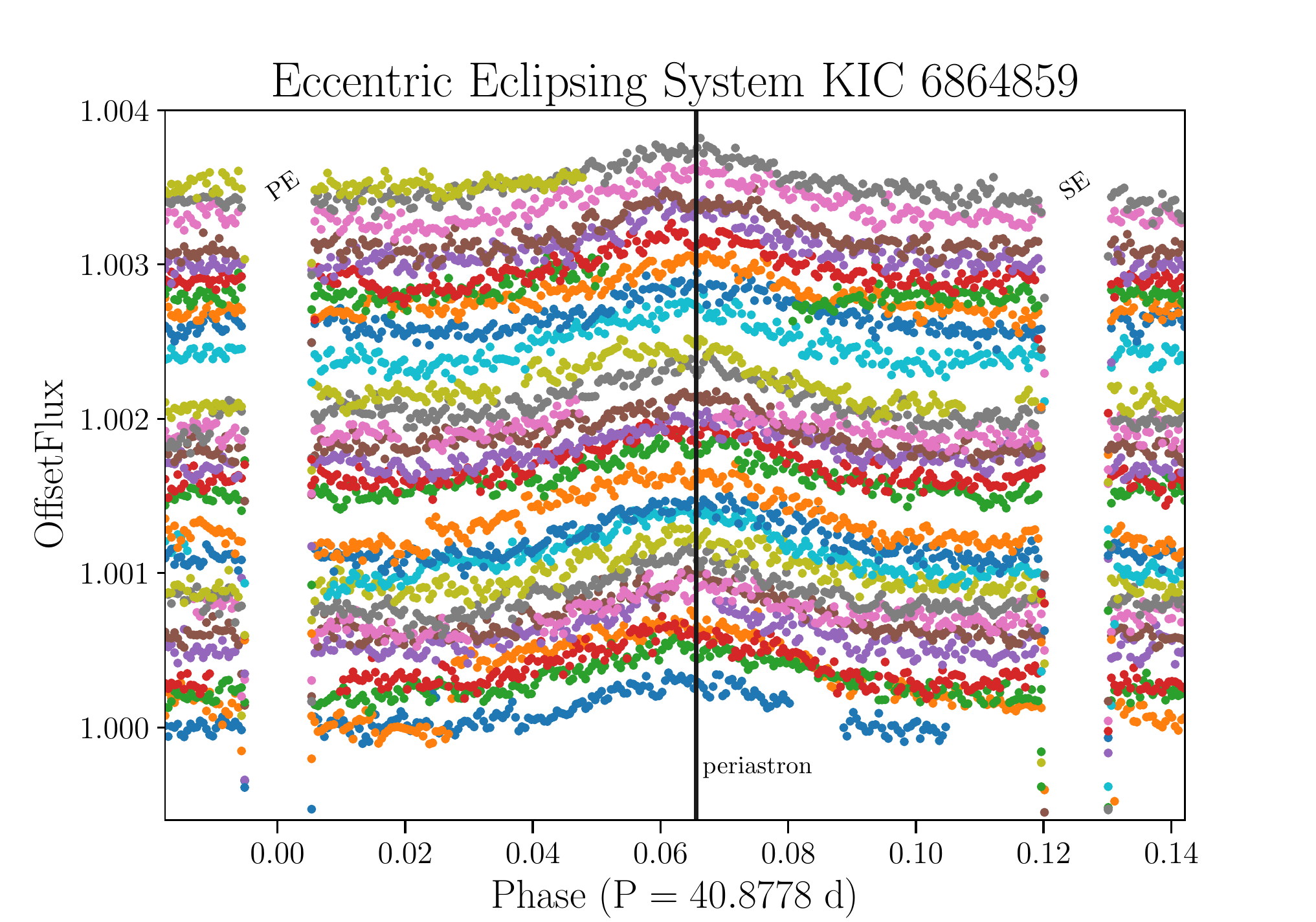}
\caption{{``Heartbeat" signatures of the eccentric eclipsing system KIC 6864659. The light curve surrounding primary and secondary eclipses is folded in phase and offset vertically between each observed orbit, using different colors for visual clarity. The brightening between primary and secondary eclipse is readily apparent around phase 0.065.
{\label{fig_heartbeat}}
}}
\end{center}
\end{figure}

Figure~\ref{fig_heartbeat} shows clear brightening events of the system between primary and secondary eclipses, with maximal amplitudes $\sim$0.3~ppt at phase $\sim$0.065, the predicted phase of periastron passage from KEBLAT. This behaviour is consistent with tidal distortions in an eccentric orbit near periastron, and is symptomatic of a class of objects known as ``heartbeat" stars \citep{thompson2012}. Both primary and secondary eclipse residuals exhibit small amplitude ($\sim$0.5~ppt) coherent structures; these are likely due to the non-spherical shape of the stars which is not explicitly modeled in KEBLAT.

We note that in our reduction process, some \texttt{apVisit} spectra were eliminated due to a very low signal-to-noise ratio that persisted after being run through our de-spike program. The majority of the remaining visits for this target resulted in well-separated peaks from the BF.

\subsection{KIC 6778289}
{\label{6778289}}

\begin{figure}[ht!]
\begin{center}
\includegraphics[width=1.1\columnwidth]{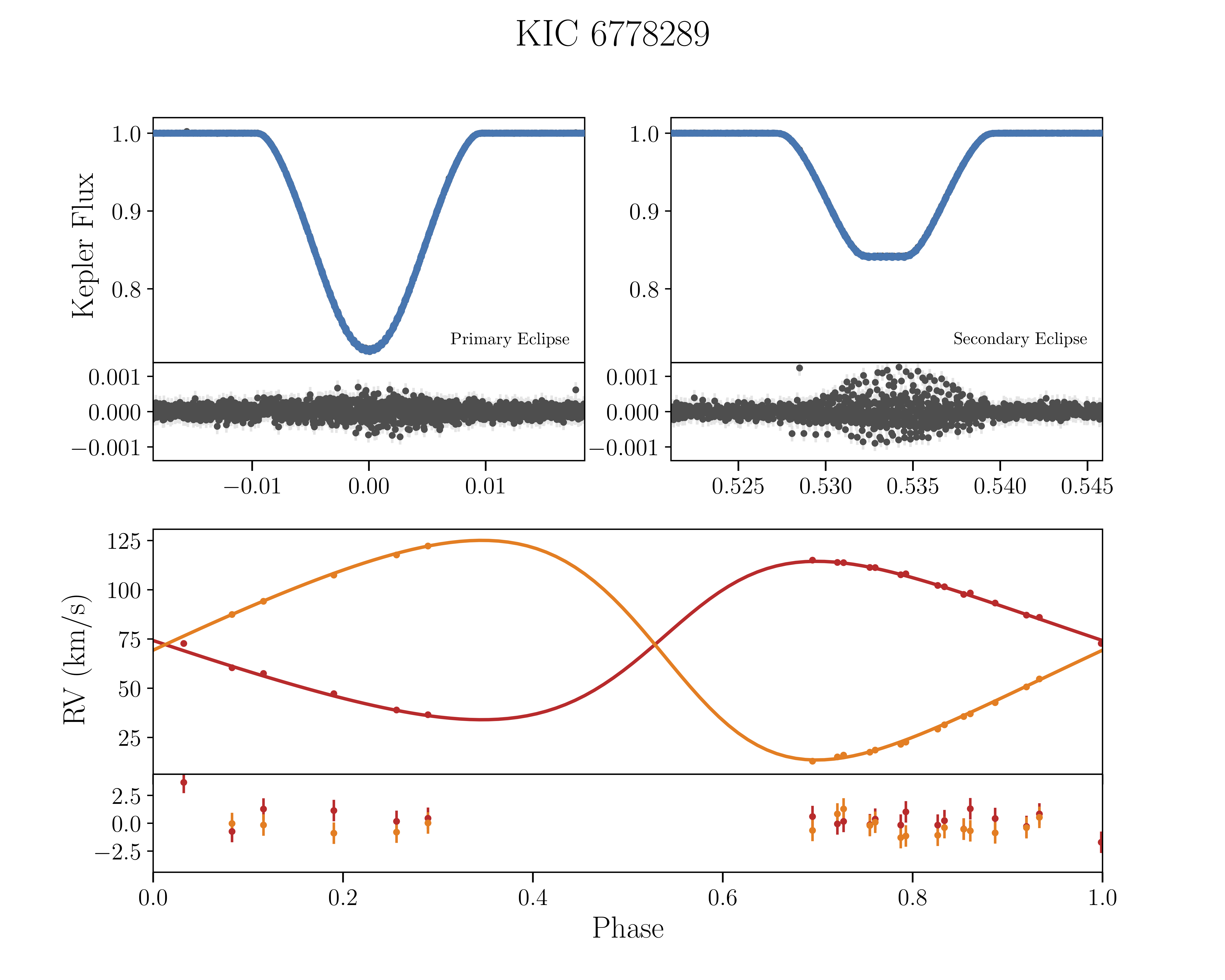}
\caption{{Simultaneous light curve (top panels) and radial velocity (bottom panel) fits to \emph{Kepler} and APOGEE data for KIC 6778289. Different eclipse depths along with a flat-bottomed secondary eclipse indicate a smaller and dimmer secondary. Additionally, the unequal amplitudes and shape of the RV indicates an unequal mass binary with significant orbital eccentricity.
{\label{fig_6778289fit}}
}}
\end{center}
\end{figure}

KIC 6778289 is a 30.1 d eclipsing binary with stellar components $M_1 = 1.51 \ M_{\odot} and M_2 = 1.09 \ M_{\odot}$ in an eccentric ($e=0.2$), nearly edge-on ($i=89.3^{\circ}$) orbit. The simultaneous RV+LC fit is shown in Figure~\ref{fig_6778289fit}. The radii are $R_1 = 1.75 \ R_{\odot}$ and $R_2 = 1.0 R_{\odot}$. This difference in radius gives rise to the difference in primary and secondary eclipse shape. The flat-bottom secondary eclipse indicates a total eclipse of the secondary component, while the primary eclipse is more V-shaped, e.g., more grazing. The larger residuals during secondary eclipse is consistent with starspot modulation. The system has low ($\approx 1 \%$) third light contamination in the \emph{Kepler} light curve which does not appear in the BF.

\subsection{KIC 6449358}
{\label{6449358}}
KIC 6449358 is a 5.8 d circular EB which may be a gravitationally bound to a distant tertiary companion. 

The BF for this object exhibits two clear peaks, however, the second-brightest peak is relatively stationary in RV while the brighter peak varies by $\sim$60 km/s over one orbit. This is shown in Figure \ref{fig_6449358fit} as well as in Appendix \ref{appendixA} Figure \ref{fig_6449358bf}. If the $\sim$stationary BF peak corresponded to a stellar binary component, it would require a system with extremely large mass ratio $M_2/M_1>10$, which would be consistent with a white dwarf. However, the light curve constrains the radius ratio to be $R_2/R_1\sim0.3$, which makes this scenario physically implausible.

A more likely explanation for the $\sim$stationary RV component is that it belongs to a tertiary star, and that the true secondary stellar component of the EB is too faint to be robustly detected by APOGEE. Indeed, the flux ratio in the \emph{Kepler} bandpass is $F_2/F_1\sim0.01$. Thus, we effectively treat KIC 6449358 as a single-lined SEB in our model, and as a result we are only able to constrain the mass function $f_M$ of the binary. Specifically, to reproduce the observed RV amplitude, our sum and ratio of masses solutions are degenerate, tending toward two extremes: high total mass ($\sim 4 \ M_{\odot}$) with a low mass ratio ($q \sim 0.3$), or low total mass ($1 \ M_{\odot}$) with a higher mass ratio ($q \sim 0.67$).

We note that some APOGEE visits do suggest a small, third BF peak (see Appendix \ref{appendixA} Figure~\ref{fig_6449358bf_forwardmodel} for details). These marginal BF peaks have large radial velocity variations from visit to visit, consistent with a low-mass star. This supports the scenario with total mass $\sim2.3 M_{\odot}$ and mass ratio $q \sim 0.45$.

\begin{figure}[!ht]
\begin{center}
\includegraphics[width=1.1\columnwidth]{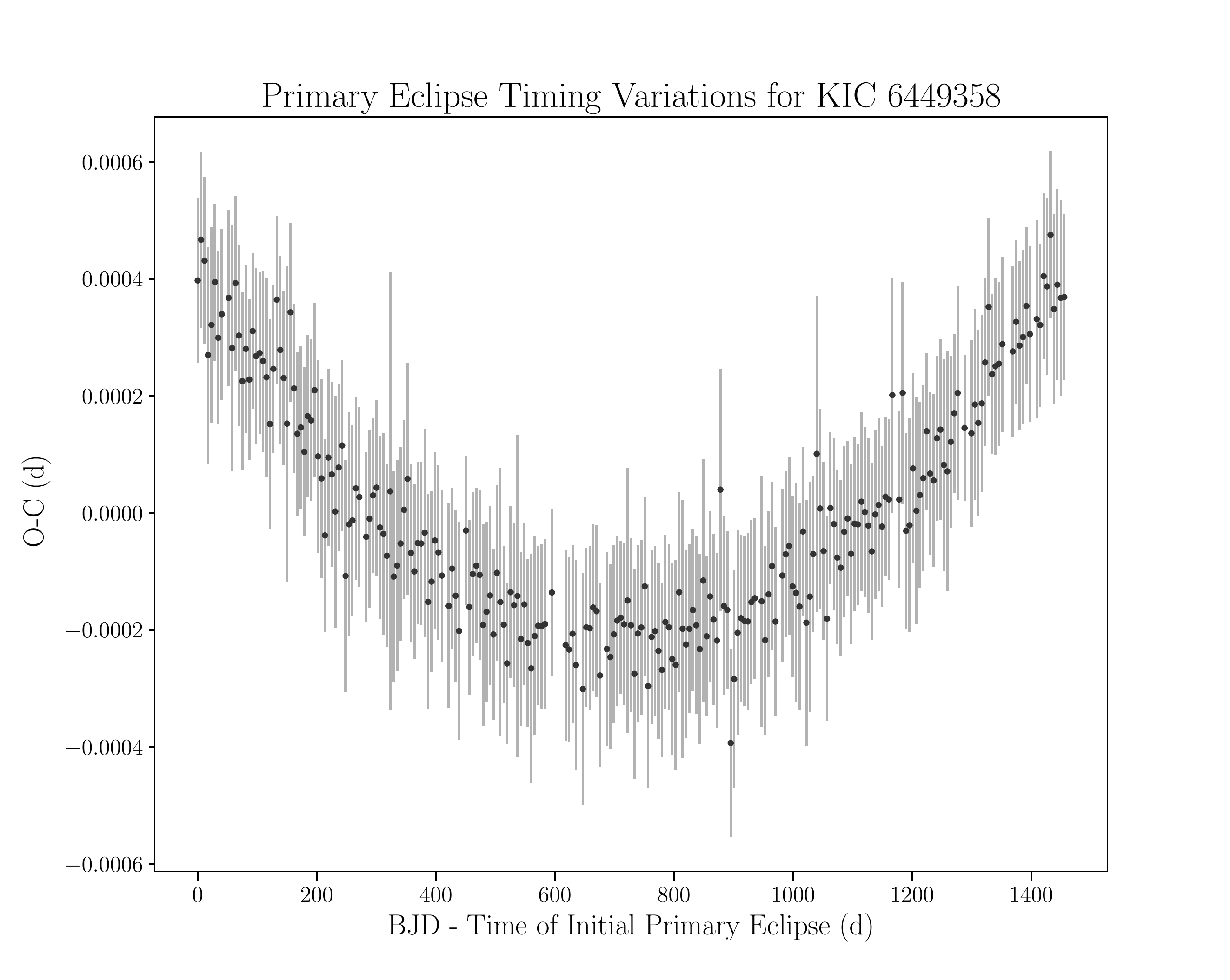}
\caption{{O-C diagram for KIC 6449358, showing primary eclipse timing variations; the ETVs exhibit a half-sinusoid trend, giving a rough estimate for the minimum period of the tertiary perturber to be $P\sim1450$ d. 
{\label{fig_6449358_oc}}
}}
\end{center}
\end{figure}

Figure~\ref{fig_6449358fit} shows the best KEBLAT model fit to the light curve and radial velocities obtained using mass function $f_M = \frac{M_2^3 \sin i}{M_1+M_2)^2}$ her than $M_1$ and $M_2$. The $\sim$stationary RV points, which are not fit, correspond to the putative third star, either a line-of-sight coincident or a tertiary companion in a hierarchical triple system. We favour the latter scenario, as the eclipses show timing variations consistent with perturbations by a bound, tertiary component. These eclipse timing variations (ETVs) have been used to identify and characterize many \emph{Kepler} EBs \citep[e.g.][]{borkovits2016}, and we fit these ETVs using a simple linear ephemeris based on the observed times of primary eclipse. We then compute the observed minus computed $(O-C)$ eclipse times as a function of time. The result is shown in Figure \ref{fig_6449358_oc}. There is a clear parabolic or sinusoidal trend in the ETVs with an amplitude of $\sim0.0006$ d; the ETV timescale indicates that the perturbing tertiary body has a minimum period $\sim1450$ d. The architecture of this type of hierarchical triple -- short, circular inner binary orbited by a distance tertiary companion -- is consistent with dynamical processing via the \cite{Kozai_1962} mechanism. 

\begin{figure}[!ht]
\begin{center}
\includegraphics[width=1.1\columnwidth]{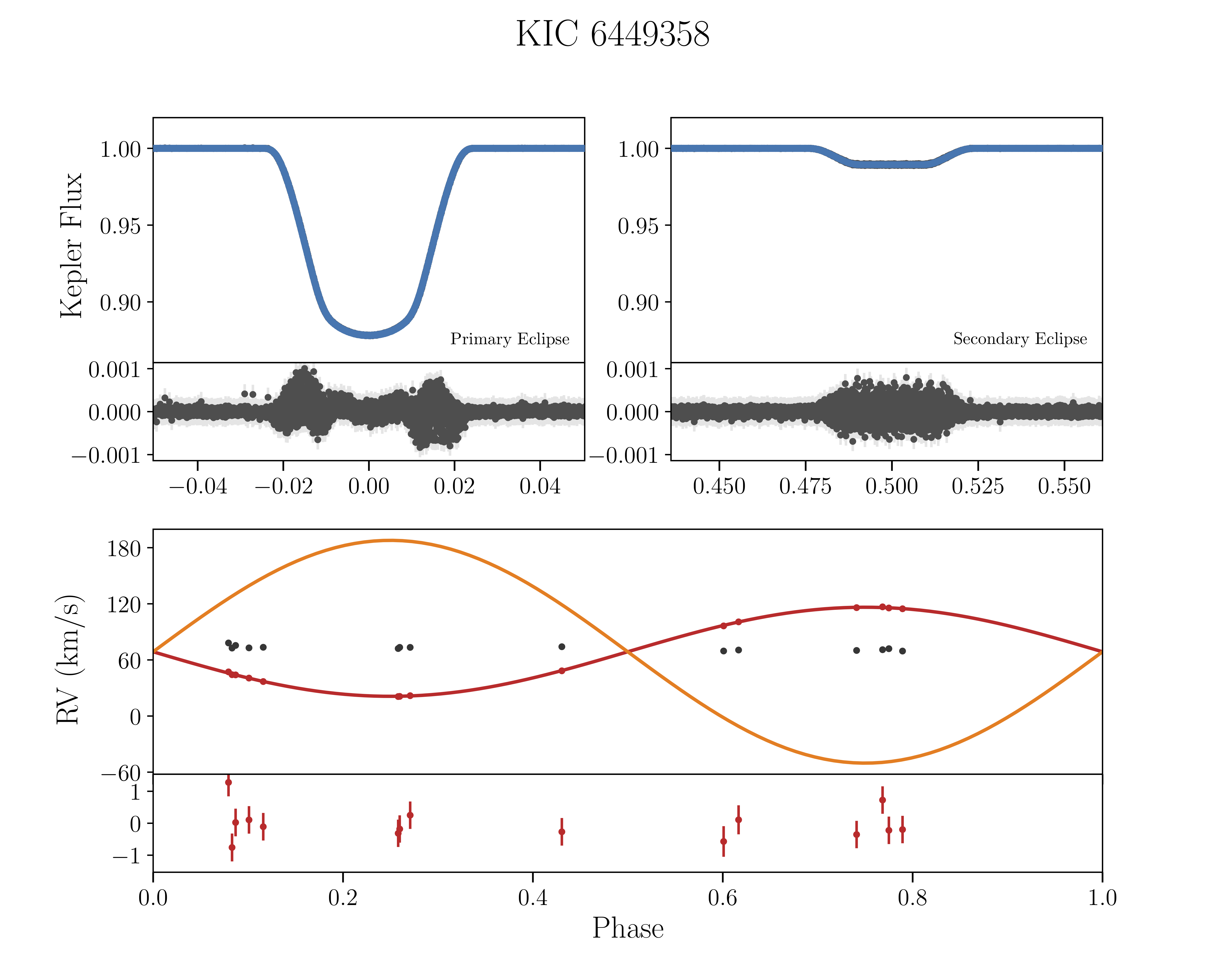}
\caption{{Simultaneous RV+LC fit for KIC 6449358, a single-lined spectroscopic binary suspected in a hierarchical triple system. We utilize mass function her than individual component masses to obtain the RV fit. The $\sim$stationary RV points near $\sim$80 km/s are measured RVs of a tertiary companion, while the orange curve shown here is the predicted radial velocity of the unseen secondary, based on the locations of tentative BF peaks in Appendix \ref{appendixA} Figure~\ref{fig_6449358bf_forwardmodel}. The flat-bottom secondary eclipse indicates the system is close to edge-on with a radius ratio $\sim0.33$, breaking the inclination-radius ratio degeneracy. The primary eclipse residuals are significant during ingress and egress, consistent with eclipse timing variations due to a tertiary companion (see Figure~\ref{fig_6449358_oc}). 
{\label{fig_6449358fit}}
}}
\end{center}
\end{figure}

\subsection{KIC 4285087}
{\label{4285087}}

\begin{figure}[ht!]
\begin{center}
\includegraphics[width=1.00\columnwidth]{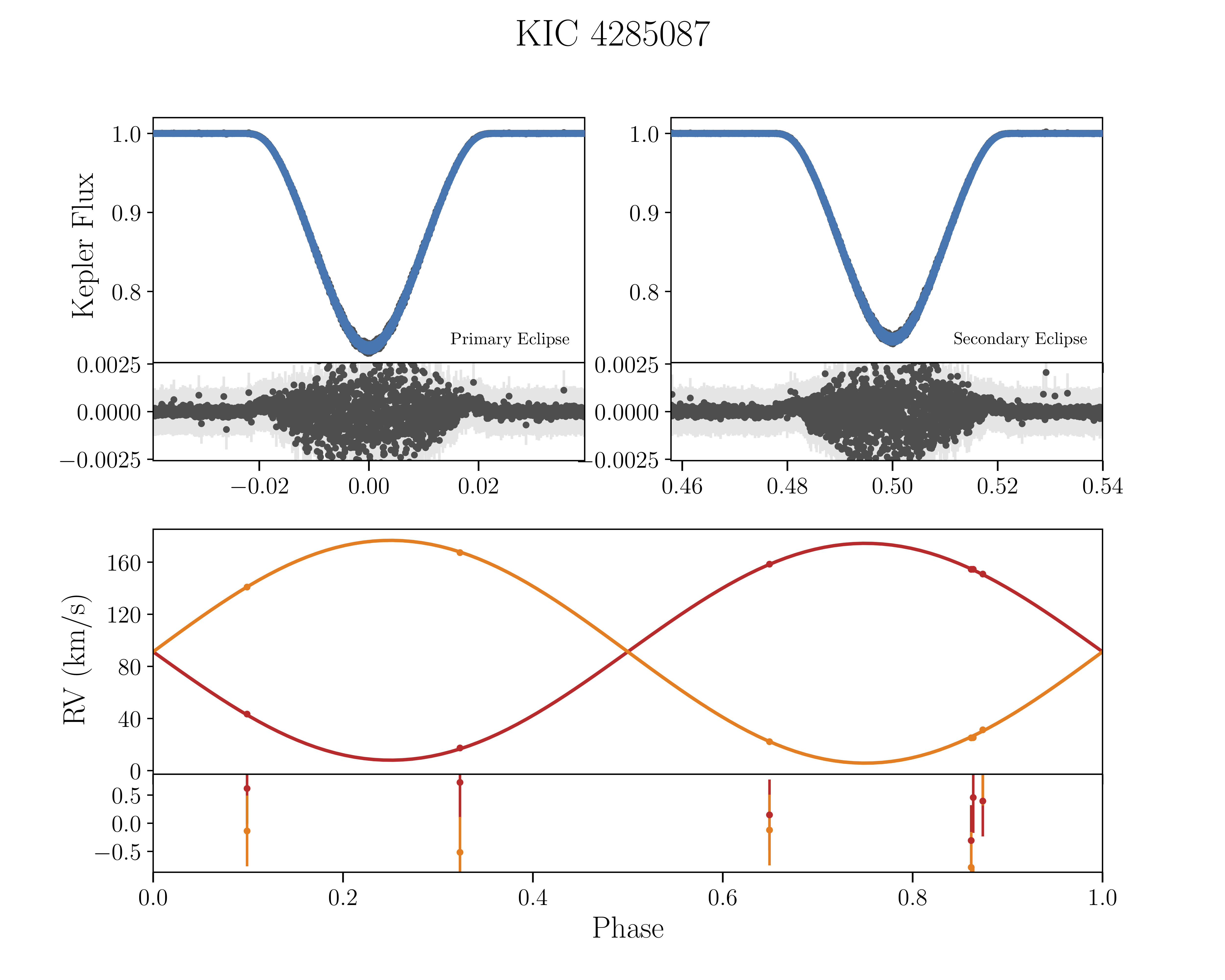}
\caption{{Simultaneous light curve (top panels) and radial velocity (bottom panel) fits to \emph{Kepler} and APOGEE observations. The phase of secondary eclipse and shape of RVs indicate EBs in a \textasciitilde{}circular orbit. The light curve residuals during eclipse suggest the presence of a variable third light contribution, or non-Keplerian photometric effects given its short period.
($\sim 4$ d).
{\label{fig_4285087fit}}
}}
\end{center}
\end{figure}

KIC 4285087 is an equal mass binary
($M_1 \approx M_2 \approx 1.1 M_{\odot}$) in a circular, slightly
inclined ($i = 87.3^{\circ}$) 4.5 d orbit. We show the best-fit solution in Figure~\ref{fig_4285087fit}. The components are
main-sequence dwarfs with similar radii
($R_1 \sim R_2 \sim 1 R_{\odot}$). The eclipses are similar depth
($\sim 30\%$), duration ($\sim 0.2$ d), and shape
(V), consistent with equal mass dwarfs orbiting each other.

\subsection{KIC 6131659}
{\label{6131659}}

\begin{figure}[!ht]
\begin{center}
\includegraphics[width=1.00\columnwidth]{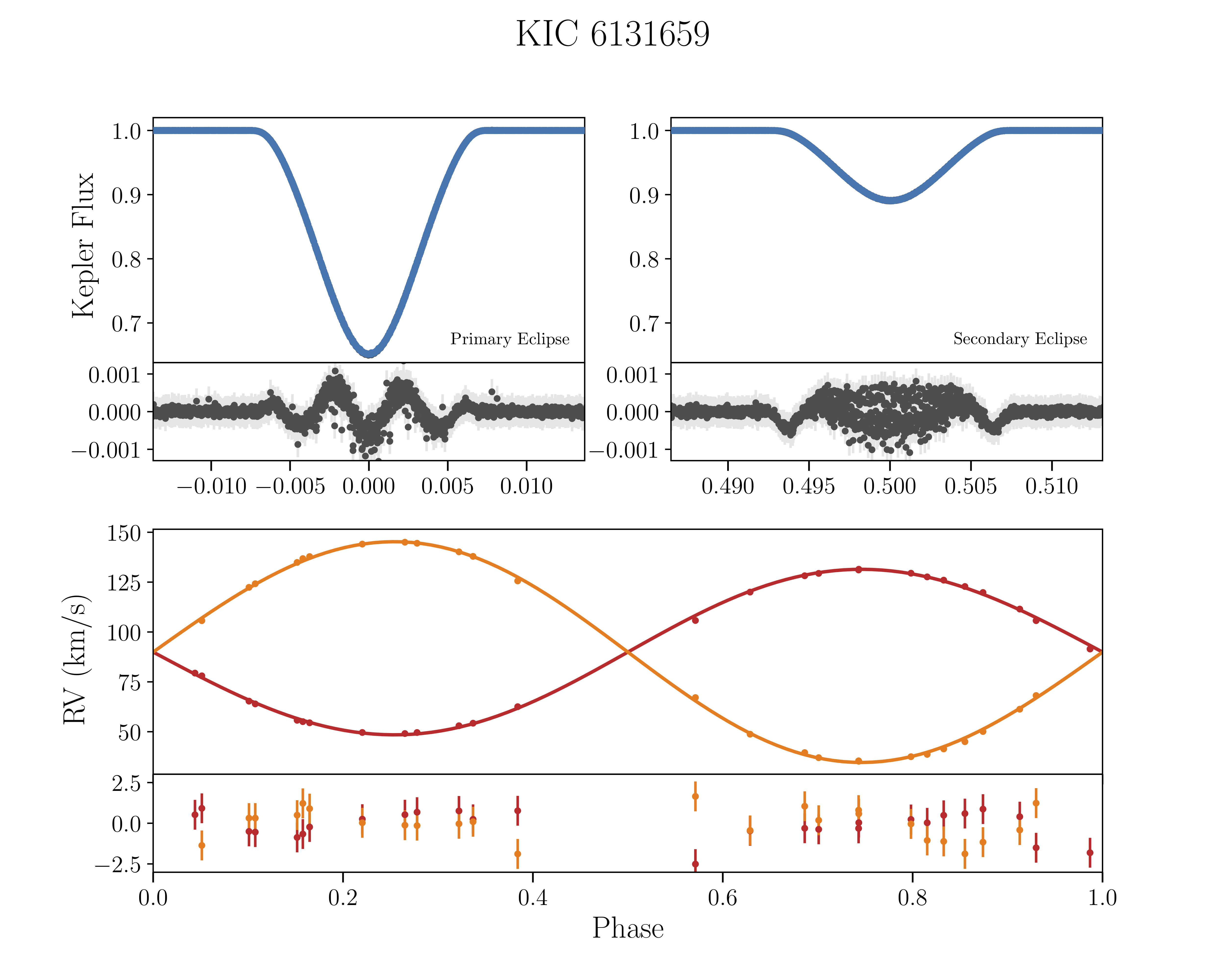}
\caption{{Simultaneous light curve (top panels) and radial velocity (bottom panel) fits to \emph{Kepler} and APOGEE observations for KIC 6131659. The primary and secondary eclipses are relatively deep, with 35\% and 10\% loss of total system light, respectively.
{\label{fig:6131659fit}}
}}
\end{center}
\end{figure}

KIC 6131659 is a mass ratio of 0.75 binary ($M_1 = 0.9 \ M_{\odot}, M_2 = 0.7 \ M_{\odot}$) in a 17.5 d, close-to-circular orbit. Figure \ref{fig:6131659fit} shows the simultaneous RV+LC fit to the data. The primary and secondary eclipses are relatively deep, with 35\% and 10\% loss of total system light, respectively. 
The residuals to the light curve fit show correlated structure, which may be due to poor limb darkening modeling and/or a varying third light component which deviates from \emph{Kepler} crowding values. 

There is a third light component readily visible in the BF (see Appendix \ref{appendixA} Figure \ref{fig_6131659bf}), but it is not
RV-variable. This suggests it may be a line-of-sight contamination source or
a gravitationally bound body in a hierarchical triple with an orbital period much longer than 17.5 d. The light curve does not show apparent eclipse timing variations, but this does not preclude the presence of a gravitationally bound tertiary.

\subsection{KIC 6781535}
{\label{6781535}}

\begin{figure}[!ht]
\begin{center}
\includegraphics[width=1.00\columnwidth]{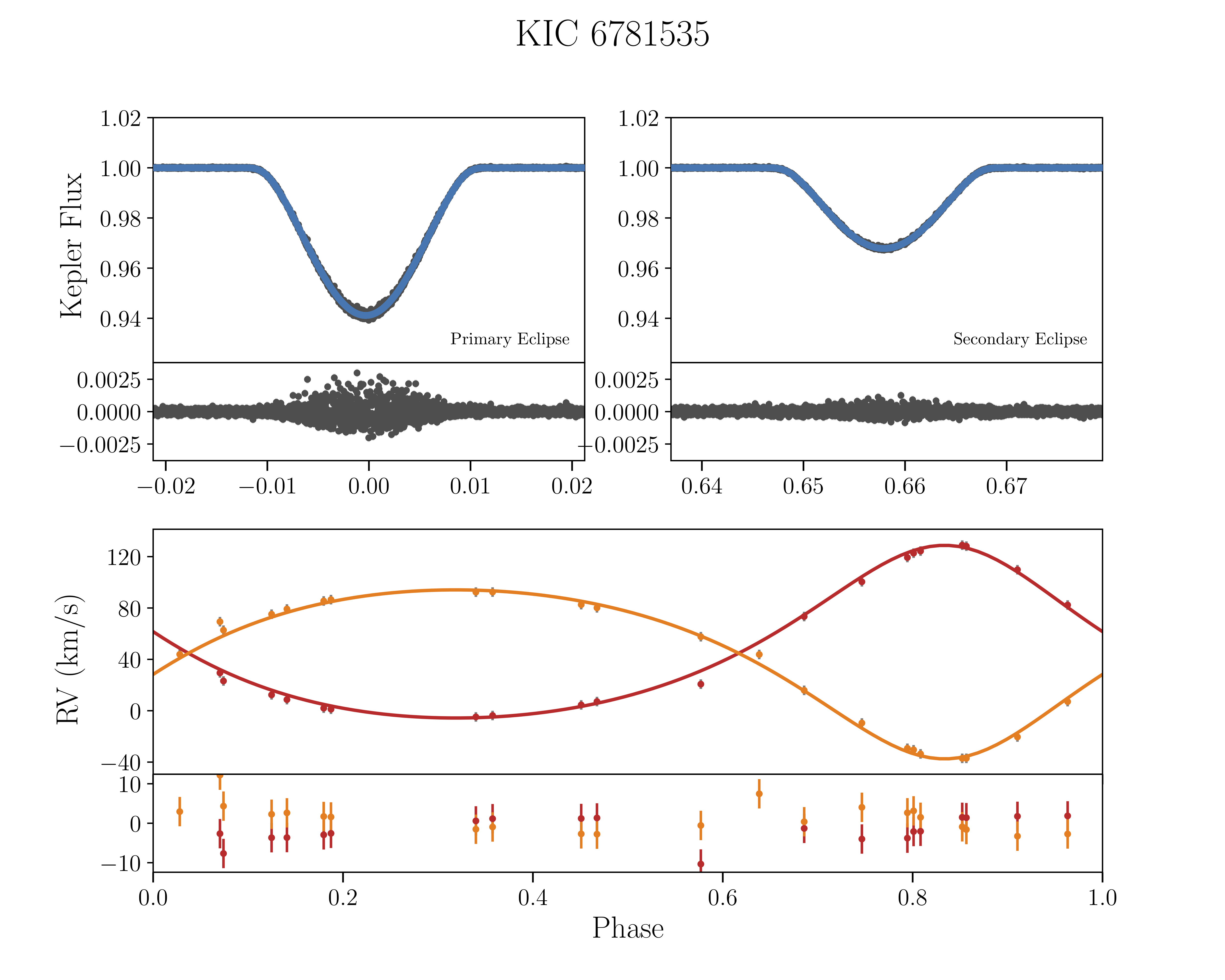}
\caption{{Top panels show the light curve model (blue) for KIC 6781535 overlaid
against data (grey) as a function of phase. Bottom panel shows
the RV model (lines) for primary (red) and secondary (orange) components
overlaid on top of APOGEE-extracted data (points). The models
correspond to the best-fit joint LC+RV solution.
{\label{fig_6781535fit}}
}}
\end{center}
\end{figure}

KIC 6781535 is an eccentric ($e=0.25$), grazing ($i=84^{\circ}$), 9.1
d binary. The best-fit solution (see Figure \ref{fig_6781535fit}) yields binary
components of similar mass $(M_2/M_1 \approx 1.0$) but slightly different radii
($R_2/R_1\approx1.2 \pm0.1$), which suggests a slightly evolved system. The shallow eclipses poorly constrain the system's impact parameter, flux ratio, and radius ratio. As a result of this degeneracy, we apply a Gaussian prior on the light curve flux ratio parameter from spectra, following the same method as used for KIC 5285607 (see \S\ref{5285607}).

Similarly to KIC 6131659, there is a third light component visible in the BF (see Figure \ref{fig_6781535bf}) that is not RV variable, indicating either a line-of-sight coincident third star or gravitationally bound tertiary companion. There are symmetric structures in the light curve residuals, most noticeably during primary eclipse, consistent with variable third light contribution, changes to the binary orbit due to additional bodies, or starspot modulations. Because the system exhibits shallow, grazing eclipses, it was not conducive to an ETV analysis. 

\subsection{Supplemental Physical Parameters}
\label{supp_parms}

In addition to the main results in Table \ref{table2}, we report some additional physical parameters in Table \ref{table3}. As discussed in Section \ref{temps}, we can use the BF peak area ratios to measure the $H$-band flux ratio of each system. We can also combine \emph{Gaia} parallax information with our measured fluxes, radii, and the ASPCAP $T_{\textrm{eff}}$ to estimate individual stellar temperatures. These parameters, along with adopted values from external sources, are summarized in Table \ref{table3}.

\input{table3.tex}  

\section{Discussion}
{\label{discuss}}

\subsection{Evolutionary Histories}
{\label{evol}}

With such well-characterized stars, we can investigate each binary's age and evolutionary history with two different approaches. In the following we assume normal Milky Way metallicities $-0.5 < $[M/H]$ < 0$.

First we explore the H-R diagram in $\log g$ vs.\ $\log T_{\rm eff}$ after first correcting our temperature estimates following \citet{ElBadry_2017}. We calculate $\log g$ for each star directly from the KEBLAT mass and radius. We also use the KEBLAT masses and radii to determine the system ages in the mass-radius space directly, which avoids any dependence on distance or on our disentangling of the individual component temperatures. 

In both approaches, we use Dartmouth evolutionary tracks \citep{Dotter_2008} and consider only the portion of the track with $\log g \ge 4.1$. This effectively only includes the main sequence. For consistency, we adhere to the KEBLAT definition of star 1 (primary) and star 2 (secondary) in which star 1 is the member of the SEB being eclipsed during the primary eclipsing event, and star 2 (secondary) as the member eclipsed during the secondary eclipsing event.

\begin{figure}[ht!]
\begin{center}
\includegraphics[width=1.1\columnwidth]{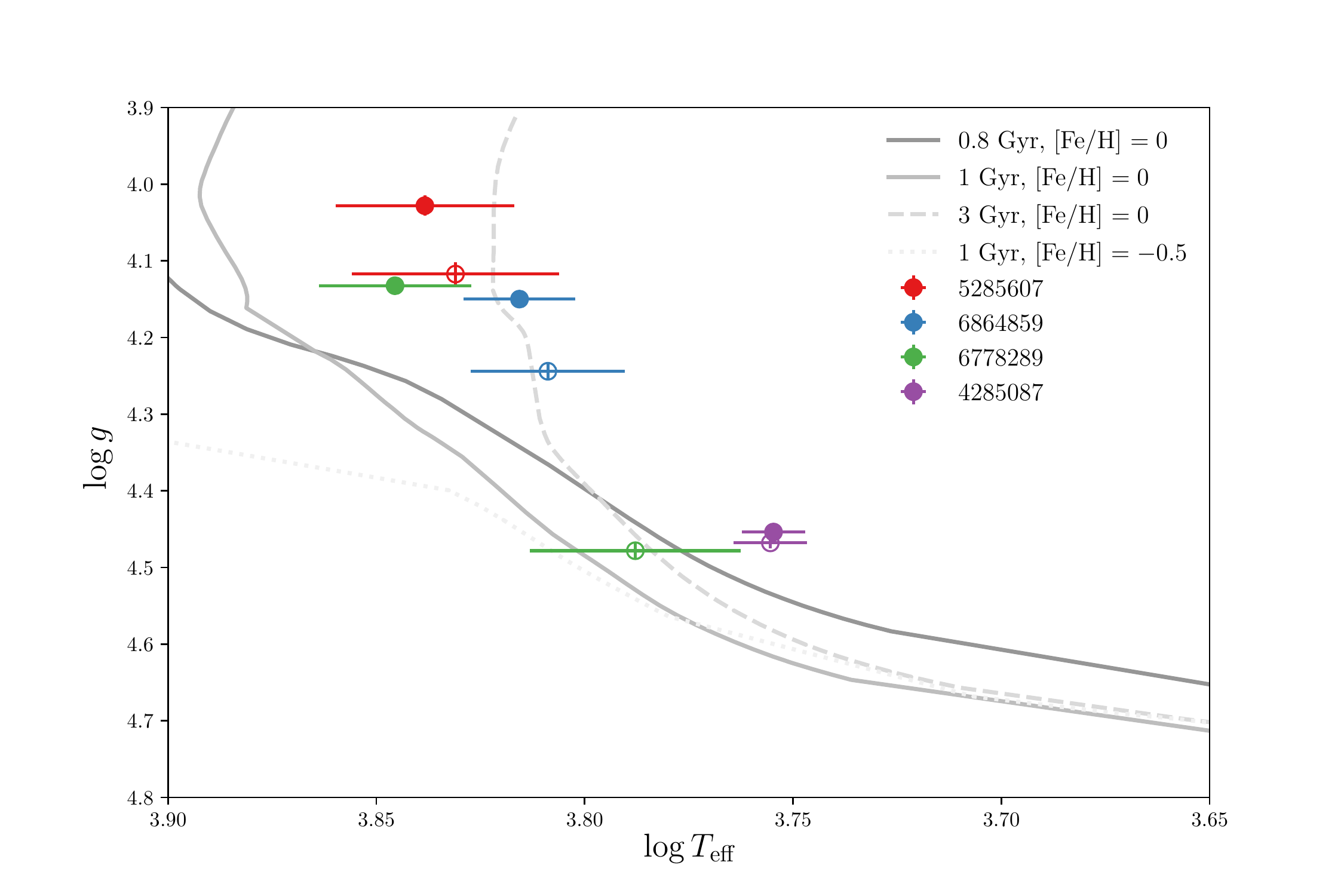}
\caption{{Spectroscopic H-R diagram for the systems with \emph{Gaia} distances. Primaries are depicted with the solid circles while secondaries are open circles. A variety of Dartmouth isochrones are plotted with a range of ages 0.8--3 Gyr and metallicities (sub-solar to solar). Only the four targets with \emph{Gaia} parallaxes are plotted here. Assuming Milky Way metallicity, we find all of our systems exhibit a high degree of coevality with ages ranging from 1--3 Gyr.
{\label{fig_loggteffall}}
}}
\end{center}
\end{figure}

\begin{figure}[ht!]
\begin{center}
\includegraphics[width=1.1\columnwidth]{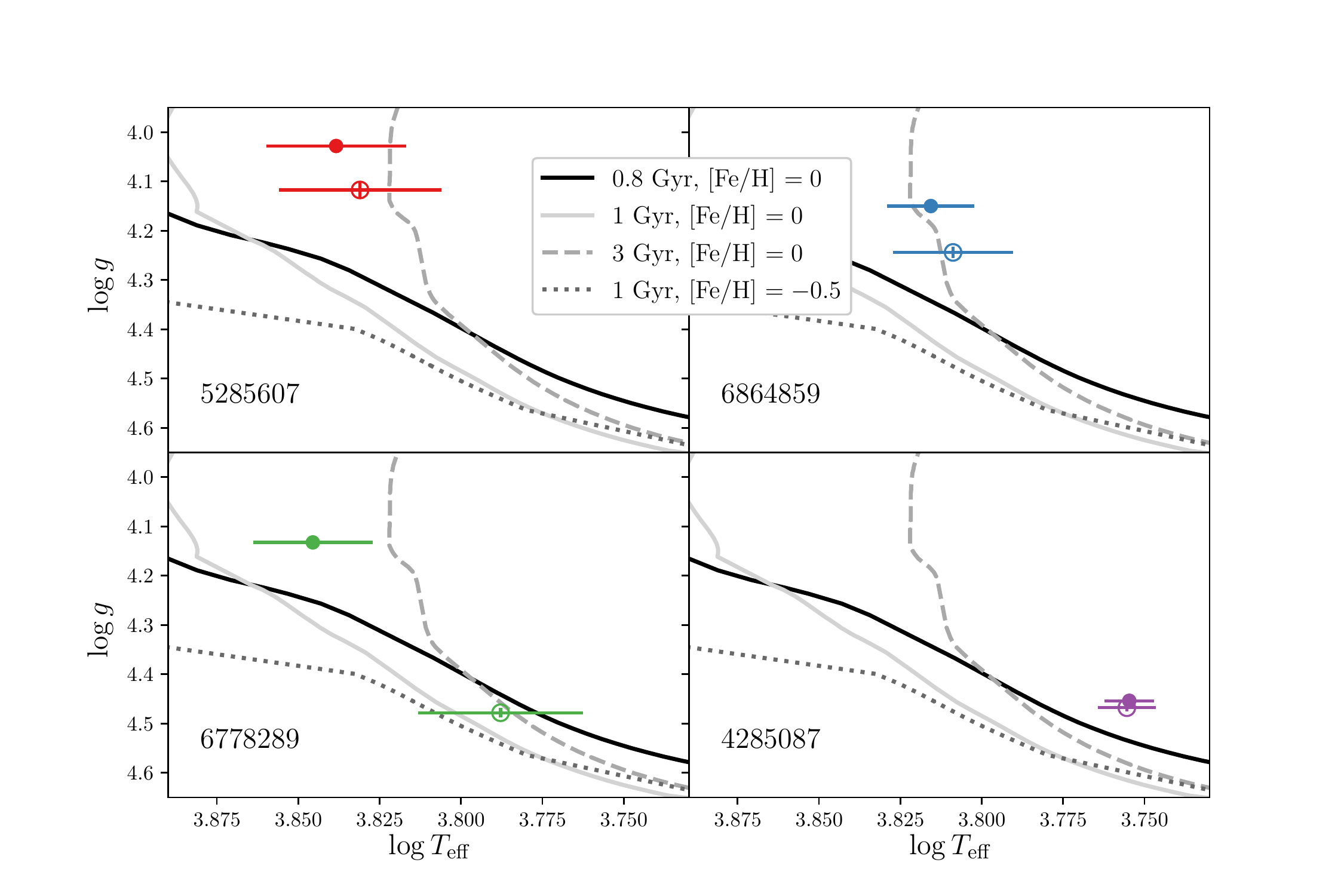}
\caption{{Figure \ref{fig_loggteffall}, showing each SEB system individually. The filled circles model the primary star of the SEB, and the hollow circles the secondaries. 
{\label{fig_loggteffsubs}}
}}
\end{center}
\end{figure}

\begin{figure}[ht!]
\begin{center}
\includegraphics[width=1.1\columnwidth]{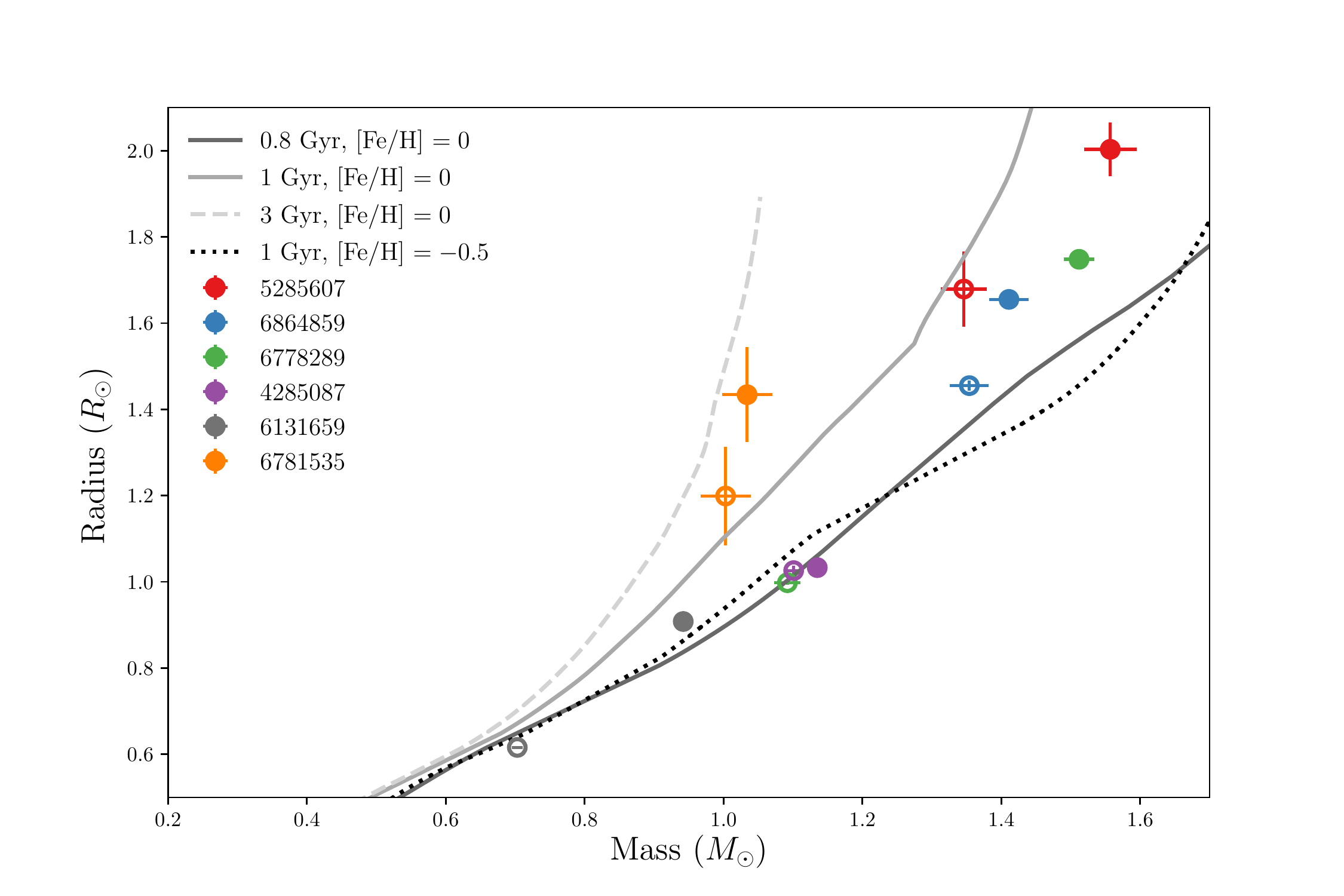}
\caption{Radius versus mass diagram for all seven systems. Primaries are shown as solid circles and secondaries in open circles. Dartmouth isochrones for a variety of ages and metallicities are also plotted. All seven systems are consistent with coevality, ranging in age from about 1 to about 3 Gyr, assuming a normal Milky Way metallicity of [M/H] $= -0.5$
{\label{fig_R_vs_M_all}}
}
\end{center}
\end{figure}

Figure~\ref{fig_loggteffall} shows all of the SEBs in the $\log g$ vs.\ $\log T_{\rm eff}$ diagram, and Figure~\ref{fig_loggteffsubs} shows each SEB system individually. In general, the systems are broadly consistent with ages ranging from about 0.8 to about 3 Gyr, and the two components of each system appear to be consistent with a common age. 
Figure~\ref{fig_R_vs_M_all} represents these systems in the mass-radius diagram, where again all six systems modeled appear consistent with coevality for the same range of ages as above. 

\subsection{Mass-Luminosity Relationships}
\begin{figure}[ht!]
\begin{center}
\includegraphics[width=1.1\columnwidth]{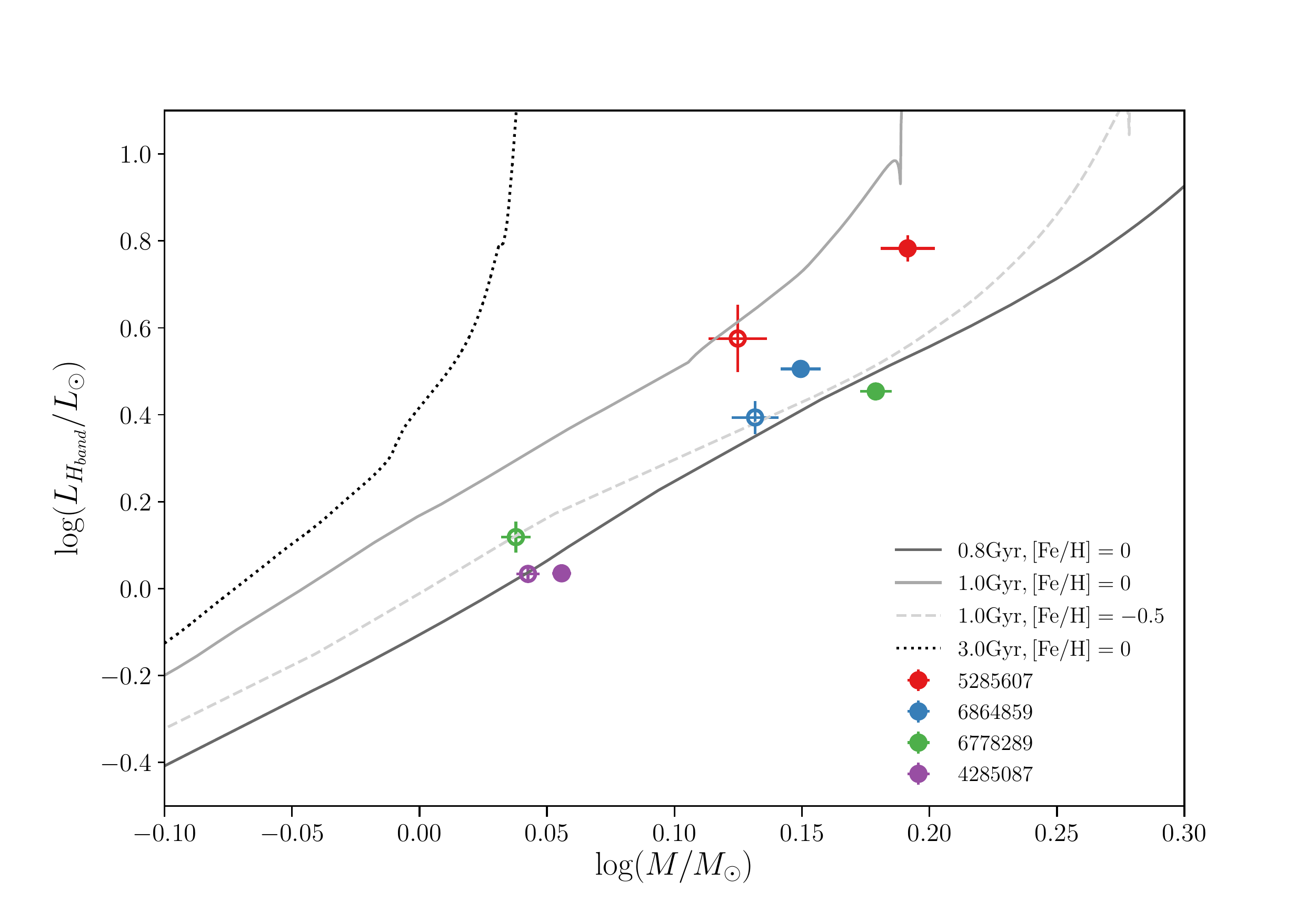}
\caption{{Mass-luminosity relationship for the four targets with accurate \emph{Gaia} distances and RV-derived masses. Primaries are shown in solid circles and secondaries in open circles. Dartmouth isochrones \citep{Dotter_2008} are overplotted as grey lines, corresponding to a variety of ages and metallicities.
{\label{fig_masslumplot}}
}}
\end{center}
\end{figure}

{\label{masslum}}
 In order to verify that our targets are on the main sequence, we create a mass-luminosity plot (Figure \ref{fig_masslumplot}) using the stellar masses from Table \ref{table2} and calculated $H$-band luminosities explained below. These luminosities are independent of the ASPCAP temperature estimates and corrections from \cite{ElBadry_2017} used to derive Figures \ref{fig_loggteffsubs} and \ref{fig_R_vs_M_all}. The results represent a comparison to theoretical models that complements the mass-radius relationship presented in Figure \ref{fig_R_vs_M_all}, and is less reliant on light curve modeling, which may have degenerate radius solutions in grazing geometries.
 
 To calculate $H$-band luminosities, we use distances derived from \emph{Gaia} parallaxes \citep{Bailer-Jones_2018} to convert apparent $H$-band magnitudes from 2MASS \citep{skrutskie2006} to absolute magnitudes. We check that the $H$-band magnitudes were not taken during eclipse by cross referencing the time of 2MASS observations to the EB ephemeris. We then compute the system $H$-band luminosities from absolute magnitudes using the sun as a reference, with $H$-band magnitude of $3.32$ from \citet{2003AJ....126.1090C}. We disentangle the separate luminosities for each stellar component in the system using the observed APOGEE $H$-band flux ratios (see Table \ref{table2}). In Figure~\ref{fig_masslumplot}, we show each system using the same plotting convention as Figures~\ref{fig_loggteffall} to \ref{fig_R_vs_M_all}, where solid and open circles correspond to primary and secondary components, respectively. We over-plot for comparison theoretical masses and $H$-band magnitudes from Dartmouth isochrones \citep{Dotter_2008} at sub-solar and solar metallicity and a range of ages (0.8--5 Gyr). In general, as previously concluded, members of the same binary system follow the same evolutionary track, i.e., are coeval.

We exclude KIC 6449398 from this analysis, because it is a single-lined SEB. We also exclude KICs 6131659 and 6781535, which have negative parallax values from \emph{Gaia}. For this reason we could not include these targets in the spectroscopic H-R diagram for our systems with \emph{Gaia} distances. We did not include reddening corrections to the distance modulus calculations; however, reddening effects should be minimal in the near-IR (2MASS and APOGEE $H$-band) compared to the visible (\emph{Kepler}). 

\subsection{Tertiary Companions}
{\label{tertiary}}

In our subset of SEBs, we identified three candidate triple
systems. The binaries with possible tertiary companions are KIC 6131659,  KIC 6781535, and KIC 6449358 (see Sections {\ref{6131659}}, 
{\ref{6781535}}, and {\ref{6449358}}, respectively). The first two of these systems
exhibit clear third BF component peaks in nearly all RV visits. We fit these third peaks with Gaussians, similar
to the Gaussian fits for the primary and secondary BF components. In both cases,
the third BF components do not have radial velocity variations above the noise, which
suggest these third members are either line-of-sight contamination
sources or gravitationally bound in hierarchical triples with orbital
periods much longer than that of the observed SEB.

While statistics on multiplicity are not complete, numerous studies have found tertiary systems composed of a tight binary orbited by a distant third member. Statistics from \cite{Tokovinin_2006} indicate that roughly 63\% of spectroscopic binaries have tertiary companions in a wide orbit. In binaries with shorter periods (less than 3 days) this percentage rises to 96\%, but in longer period binaries (12 or more days) this percentage is only 34\%. 

The incidence of triples among \emph{Kepler} close eclipsing binaries (as are the ones in our analysis) is at least $\sim$20\% \citep{rappaport2013, conroy2014}, and likely higher for tertiaries with longer periods beyond \emph{Kepler}'s finite observing time.

Evidence from both spectroscopic and photometric observations indicate KIC 6449358 belongs to a hierarchical triple system (see \S \ref{6449358}). The BF for this system shows a stationary tertiary peak in a few of the APOGEE visits.  The mid times of eclipses in the \emph{Kepler} light curve also exhibit sinusoidal variations in time; these eclipse timing variations indicate that the tertiary has a period $\gtrsim1450$~d.

Interestingly, among our sample we do not detect tertiary companions among the shortest-period binaries. In particular, neither of the two EBs with $P_{\rm orb} < 5$~d exhibits a clear tertiary in our data. It is not yet clear whether existing observations might already exclude the presence of tertiaries at very large separations that might not appear in our data; additional imaging observations might be required to identify such stars. At the same time, two of the four EBs with $P_{\rm orb} > 9$~d are triples, which would appear to be an over-abundance of tertiaries among the longest-period EBs, albeit with a small sample. However, we note that one of these (KIC 6781535, $P_{\rm orb} \approx 9$~d) is a modestly evolved system (see Figure~\ref{fig_R_vs_M_all}), and excluding that case yields an occurrence of 1/3 triples among our EBs with $P_{\rm orb} > 12$~d), fully consistent with the results of \citet{Tokovinin_2006}.

\section{Summary}
{\label{summary}}
We thoroughly characterize seven SEBs that have been observed by both \emph{Kepler} and APOGEE. Our targets are selected from the \emph{Kepler} EB catalog, and limited to bright, detached EB targets with both primary and secondary eclipses observed by \emph{Kepler}, high inclination, and multiple APOGEE visits. We identify an additional 26 SEBs which may warrant similar studies. We demonstrate that the BF is a superior method for extracting RVs from APOGEE visit spectra compared to the CCF used in the present reduction pipeline. This is particularly true for systems with multiple RV variable components. While such an analysis is beyond the scope of this work, if the BF method were applied to the full data set of multiple-visit APOGEE targets, it would most likely reveal many previously unknown SEBs and other interesting RV-variable sources.

RVs are extracted from \texttt{apVisit} spectra using the BF method and the \emph{Kepler} light curves are normalized, de-weighted and modeled using KEBLAT. The light curve and RV solutions are first determined individually, and then computed simultaneously. We use the resulting physical parameters to estimate stellar temperatures, investigate coevality, and explore candidate triple systems.

Using our analysis we find our target's binary members are coeval with ages ranging from 1 to 3 Gyr, assuming normal Milky Way metallicity $(-0.5 < [M/H] < 0)$. The exception is KIC 6781535 which lies closer to a slightly metal poor $([M/H] \sim -1.0)$ 3 Gyr isochrone. Our systems being broadly consistent with coevality confirms a common assumption in star formation that members of multiples form at the same time, and also effectively calibrates stellar evolution modeling.

Overlap between large scale surveys like APOGEE, \emph{Kepler}, and \emph{Gaia} allows us to discover and analyze many diverse SEBs, including systems with very low flux ratios and those in higher order systems. The statistics on the triples within our subset with respect to the orbital period of inner binaries is broadly consistent with statistics from the field \cite{Tokovinin:1997}, though there may be some tension with our sample in that the shortest-period EBs do not appear to be spectroscopic triples. This is in contrast to the expectation that shortest-period EBs are most likely to be hierarchical triples. It is possible that very wide tertiaries do exist in these systems but have yet to be identified via imaging.

We have shown that through tools like KEBLAT and the BF analysis of APOGEE spectra, it is possible to perform high quality analysis of large numbers of SEBs with a variety of properties. This opens up great promise for future SEBs identified in TESS and SDSS-V data.

\acknowledgements
We would like to thank Scott Fleming for critical guidance and discussion and Paul A. Mason for valuable brainstorming and advice. We recognize the SDSS Faculty And Student Team (FAST) initiative for supporting this work through a partnership with New Mexico State University. JMCC thanks the Fisk-Vanderbilt Masters-to-PhD Bridge Program, Amanda Cobb, Kelly Holley-Bockelmann, and Nancy Chanover for continued empowerment of a woman and new mother in STEM. MLR celebrates that this work has encompassed two births, one wedding, and multiple graduations among the lead authors.

\software{Astropy \citep{astropy:2013, astropy:2018},
          PyAstronomy \citep{Czesla},
          Matplotlib \citep{Hunter:2007},
          Apogee \citep{Bovy_2016},
          Makecite \citep{makecite:2018},
          Numpy \citep{numpy:2011},
          Pandas \citep{pandas:2010},
          Emcee \citep{emcee:2013},
          Gaussfitter \citep{gaussfitter},
          Scipy \citep{scipy:2001}
          }

\appendix

\input{appendixA.tex}

\newpage

\bibliographystyle{aasjournal}
\bibliography{biblio.bib}

\end{document}

%% file: table1.tex
\setcounter{table}{1}
\begin{deluxetable*}{rrrcrccll}
\tablecolumns{9}
\tablewidth{0pt}
\tabletypesize{\scriptsize}
\tablecaption{Promising SEBs observed by APOGEE and Kepler, sorted by Kepler magnitude ($K_p$)}
\label{table1}
\centering
\tablehead{
\colhead{KIC} & \colhead{APOGEE ID} & \colhead{Visits} & \colhead{$K_p$} & \colhead{$P_{\textrm{orb}}$ (day)} & \colhead{SE Depth (frac)} & \colhead{Morphology} & \colhead{Reference} & \colhead{Notes} 
}
\decimals
\startdata
9246715  & 2M20034832+4536148  &  2 & 10.08   & 171.28   & 0.1124    & 0.11 & \citet{Rawls_2016} &   \\
2708156  & 2M19302686+4318185  &  3 & 10.67   &   1.89   & 0.0625    & 0.57 &  & Only 3 visits \\
3120320  & 2M19291007+3817041  &  3 & 11.28   &  10.27   & 0.0127    & 0.14 &  & Kepler APOGEE/EB WG  \\
4851217  & 2M19432016+3957081  &  6 & 11.32   &   2.47   & 0.1815    & 0.58 &  & Low S/N ratio \\
3439031  & 2M19203184+3830492  &  3 & 11.50   &   5.95   & 0.4156    & 0.33 &  & Kepler APOGEE/EB WG  \\
5285607  & 2M19390532+4027346  &  6 & 11.69   &   3.90   & 0.0403    & 0.36 & This work & \\
6449358  & 2M19353513+4149543  & 25 & 11.72   &   5.78   & 0.0120    & 0.31 & This work & \\
10206340 & 2M19245882+4714573  &  3 & 11.78   &   4.56   & 0.2431    & 0.61 &  & Only 3 visits \\
6864859  & 2M19292405+4223363  & 25 & 11.93   &  40.88   & 0.2426    & 0.06 & This work & \\
4931073  & 2M19351913+4001522  &  6 & 12.18   &  26.95   & 0.0564    & 0.08 &  & Kepler APOGEE/EB WG \\
3127817  & 2M19355993+3813561  &  6 & 12.24   &   4.33   & 0.0512    & 0.48 &  & Kepler APOGEE/EB WG \\
3335816  & 2M19184759+3824238  &  3 & 12.40   &   7.42   & 0.0106    & 0.16 &  & Kepler APOGEE/EB WG \\
5284133  & 2M19373173+4027078  &  6 & 12.50   &   8.78   & 0.0492    & 0.15 &  & Future work \\
3542573  & 2M19232622+3838017  &  3 & 12.61   &   6.94   & 0.0837    & 0.25 &  & Only 3 visits \\
2711114  & 2M19240341+3758109  &  3 & 12.63   &   2.86   & 0.0022    & 0.29 &  & Only 3 visits \\
4281895  & 2M19441242+3923418  &  6 & 12.76   &   9.54   & 0.0652    & 0.13 &  & Only 3 visits \\
4660997  & 2M19340328+3942410  &  6 & 12.78   &   0.56   & 0.2527    & 0.62 &  & Ellipsoidal variations \\
4473933  & 2M19363898+3933105  &  6 & 12.87   & 103.59   & 0.0126    & 0.25 &  & Low S/N ratio \\
2305543  & 2M19280644+3736023  &  3 & 12.97   &   1.36   & 0.1052    & 0.50 &  & Only 3 visits \\
3241619  & 2M19322278+3821405  &  3 & 13.06   &   1.70   & 0.1625    & 0.44 &  & Only 3 visits \\
4285087  & 2M19463571+3919069  &  6 & 13.19   &   4.49   & 0.2408    & 0.31 & This work &  \\
2576692  & 2M19263432+3748513  &  3 & 13.19   &  87.88   & 0.2588    & 0.04 &  & Kepler APOGEE/EB WG \\
6131659  & 2M19370697+4126128  & 27 & 13.20   &  17.53   & 0.1036    & 0.09 & This work & \\
4847832  & 2M19401839+3957298  &  6 & 13.20   &  30.96   & 0.3200    & 0.08 &  & Kepler APOGEE/EB WG \\
5025294  & 2M19414825+4010323  &  6 & 13.27   &   5.46   & 0.0010    & 0.18 &  & Future work \\
6778289  & 2M19282456+4215080  & 25 & 13.31   &  30.13   & 0.1619    & 0.11 & This work & \\
3248332  & 2M19383951+3819588  &  6 & 13.37   &   7.36   & 0.0974    & 0.20 &  & Kepler APOGEE/EB WG \\
6610219  & 2M19320615+4200049  & 25 & 13.58   &  11.30   & 0.2899    & 0.20 &  & Low S/N ratio \\
6781535  & 2M19321788+4216489  & 25 & 14.14   &   9.12   & 0.0305    & 0.12 & This work & \\
4077442  & 2M19452193+3908260  &  6 & 14.35   &   0.69   & 0.0703    & 0.59 &  & Kepler APOGEE/EB WG \\
3247294  & 2M19374558+3822510  &  6 & 14.35   &  67.42   & 0.1032    & 0.02 &  & Kepler APOGEE/EB WG\\
3848919  & 2M19241352+3858278  &  3 & 14.48   &   1.05   & 0.3418    & 0.57 &  & Low S/N ratio \\
5460835  & 2M19411125+4039416  &  6 & 14.72   &  21.54   & 0.0228    & 0.06 &  & Low S/N ratio \\
4075064  & 2M19432862+3908535  &  6 & 15.71   &  61.42   & 0.0821    & 0.00 &  & Low S/N ratio \\
\enddata
\end{deluxetable*}

%% file: table2.tex
\begin{rotatetable*}
\begin{deluxetable*}{lccccccc}
\tablecolumns{8}
\tablecaption{Binary Orbital and Stellar Parameters}
\label{table2}
\centering
\tablenum{2}
\tablehead{
\colhead{} & \colhead{KIC 5285607\tablenotemark{a}} & \colhead{KIC 6864859} & \colhead{KIC 6778289} & \colhead{KIC 6449358} & \colhead{KIC 4285087} & \colhead{KIC 6131659} & \colhead{KIC 6781535\tablenotemark{a}}
}
\startdata
\vspace{0.1cm}                                                    
$f_M$\tablenotemark{b}                                            & \nodata                                                                       & \nodata                                                             & \nodata                                                           & $0.126 ^{+0.002}_{-0.002}$                                      & \nodata                                                                       & \nodata                                                                     & \nodata                                                            \\ \vspace{0.1cm} 
$M_1~(M_{\odot})$                                                 & $ 1.557 ^{+ 0.038 }_{ -0.035 }$                                               & $ 1.411 ^{+ 0.028 }_{ -0.028 }$                                     & $ 1.512 ^{+ 0.022 }_{ -0.022 }$                                   & \nodata                                                         & $ 1.135 ^{+ 0.013 }_{ -0.014 }$                                               & $ 0.942 ^{+ 0.010 }_{ -0.010 }$                                             & $ 1.003 ^{+ 0.039 }_{ -0.038 }$                                    \\ \vspace{0.1cm} 
$M_2~(M_{\odot})$                                                 & $ 1.346 ^{+ 0.033 }_{ -0.033 }$                                               & $ 1.354 ^{+ 0.028 }_{ -0.028 }$                                     & $ 1.092 ^{+ 0.019 }_{ -0.018 }$                                   & \nodata                                                         & $ 1.101 ^{+ 0.013 }_{ -0.014 }$                                               & $ 0.703 ^{+ 0.008 }_{ -0.008 }$                                             & $ 1.034 ^{+ 0.040 }_{ -0.040 }$                                    \\ \vspace{0.1cm} 
$R_1~(R_{\odot})$                                                 & $ 2.003 ^{+ 0.062 }_{ -0.054 }$                                               & $ 1.655 ^{+ 0.012 }_{ -0.013 }$                                     & $ 1.748 ^{+ 0.009 }_{ -0.009 }$                                   & $2.1254 ^{+0.0007}_{-0.0006}$                                   & $ 1.033 ^{+ 0.010 }_{ -0.012 }$                                               & $ 0.908 ^{+ 0.003 }_{ -0.003 }$                                             & $ 1.199 ^{+ 0.113 }_{ -0.065 }$                                    \\ \vspace{0.1cm} 
$R_2~(R_{\odot})$                                                 & $ 1.679 ^{+ 0.063 }_{ -0.087 }$                                               & $ 1.455 ^{+ 0.012 }_{ -0.012 }$                                     & $ 0.998 ^{+ 0.005 }_{ -0.005 }$                                   & $0.6977 ^{+0.0005}_{-0.0004}$                                   & $ 1.026 ^{+ 0.011 }_{ -0.011 }$                                               & $ 0.616 ^{+ 0.003 }_{ -0.003 }$                                             & $ 1.434 ^{+ 0.066 }_{ -0.114 }$                                    \\ \vspace{0.1cm} 
$P_{\textrm{orb}}~(\textrm{d})$                                   & $ 3.89940111 ^{+ 4.4\mathrm{e}\minus8 }_{ -4.5\mathrm{e}\minus8 }$            & $ 40.8778425 ^{+ 2.9\mathrm{e}\minus7 }_{ -3.0\mathrm{e}\minus7 }$  & $ 30.1301385 ^{+ 3.5\mathrm{e}\minus7}_{ -3.5\mathrm{e}\minus7 }$ & $5.77679431 ^{+5.0\mathrm{e}\minus8}_{-4.9\mathrm{e}\minus8}$   & $ 4.48603142 ^{+ 5.1\mathrm{e}\minus8 }_{ -5.4\mathrm{e}\minus8 }$            & $ 17.52782741 ^{+ 2.2\mathrm{e}\minus7 }_{ -2.3\mathrm{e}\minus7 }$         & $ 9.12208640 ^{+ 2.5\mathrm{e}\minus7 }_{ -2.4\mathrm{e}\minus7 }$ \\ \vspace{0.1cm} 
$t_{pe}~(\textrm{d})$\tablenotemark{c}                            & $ 126.577647 ^{+ 1.1\mathrm{e}\minus5 }_{ -9.7\mathrm{e}\minus6 }$            & $ 158.3189733 ^{+ 9.2\mathrm{e}\minus6 }_{ -7.6\mathrm{e}\minus6 }$ & $ 122.100819 ^{+ 1.0\mathrm{e}\minus5}_{ -1.0\mathrm{e}\minus5 }$ & $122.0739252 ^{+7.1\mathrm{e}\minus6}_{ -7.2\mathrm{e}\minus6}$ & $ 133.450881 ^{+ 1.1\mathrm{e}\minus5 }_{ -1.1\mathrm{e}\minus5 }$            & $ 127.042190 ^{+ 1.1\mathrm{e}\minus5 }_{ -1.1\mathrm{e}\minus5 }$          & $ 138.838193 ^{+ 4.8\mathrm{e}\minus5 }_{ -6.6\mathrm{e}\minus5 }$ \\ \vspace{0.1cm} 
$i~(\textrm{rad})$                                                & $ 1.3656 ^{+ 0.0015 }_{ -0.0009 }$                                            & $ 1.5415 ^{+ 0.0001 }_{ -0.0001 }$                                  & $ 1.55876 ^{+ 0.00001 }_{ -0.00001 }$                             & $1.5327^{+0.0001}_{ -0.0001}$                                   & $ 1.5210 ^{+ 0.0001 }_{ -0.0001 }$                                            & $ 1.5558 ^{+ 0.0001 }_{ -0.0001 }$                                          & $ 1.4749 ^{+ 0.0007 }_{ -0.0004 }$                                 \\ \vspace{0.1cm} 
$e\sin\omega$                                                     & $ 1.3\mathrm{e}\minus6 ^{+ 9.0\mathrm{e}\minus4 }_{ -1.2\mathrm{e}\minus4 }$  & $ -0.0254 ^{+ 0.0006 }_{ -0.0005 }$                                 & $ -0.2337 ^{+ 0.0004 }_{ -0.0004 }$                               & $0.00060^{ +0.00046}_{-0.00040}$                                & $ 1.8\mathrm{e}\minus4 ^{+ 1.4\mathrm{e}\minus3 }_{ -3.0\mathrm{e}\minus4 }$  & $ -0.0127 ^{+ 0.0004 }_{ -0.0005 }$                                         & $ 0.0215 ^{+ 0.0034 }_{ -0.0026 }$                                 \\ \vspace{0.1cm} 
$e\cos\omega$                                                     & $ -8.6\mathrm{e}\minus7 ^{+ 3.5\mathrm{e}\minus6 }_{ -4.1\mathrm{e}\minus6 }$ & $ -0.634115 ^{+ 0.00001 }_{ -0.00001 }$                             & $ 0.051090 ^{+ 0.000005 }_{ -0.000005 }$                          & $-0.00023^{ +1.0\mathrm{e}\minus5}_{ -1.1\mathrm{e}\minus5}$    & $ -5.9\mathrm{e}\minus5 ^{+ 3.7\mathrm{e}\minus6 }_{ -3.8\mathrm{e}\minus6 }$ & $ 7.1\mathrm{e}\minus5 ^{+ 1.7\mathrm{e}\minus6 }_{ -1.8\mathrm{e}\minus6 }$& $ 0.25005 ^{+ 0.00001 }_{ -0.00001 }$                              \\ \vspace{0.1cm} 
$F_2/F_1$                                                         & $ 0.6579 ^{+ 0.0959 }_{ -0.1025 }$                                            & $ 0.7256 ^{+ 0.0139 }_{ -0.0124 }$                                  & $ 0.19155 ^{+ 0.00002 }_{ -0.00002 }$                             & $0.010752^{ +6.5\mathrm{e}\minus6}_{-6.5\mathrm{e}\minus6}$     & $ 0.9455 ^{+ 0.0396 }_{ -0.0356 }$                                            & $ 0.1480 ^{+ 0.0010 }_{ -0.0010 }$                                          & $ 1.0654 ^{+ 0.2195 }_{ -0.2384 }$                                 \\ \vspace{0.1cm} 
$q_{1,1}$                                                         & $ 0.3690 ^{+ 0.0158 }_{ -0.0157 }$                                            & $ 0.4313 ^{+ 0.0158 }_{ -0.0176 }$                                  & $ 0.2930 ^{+ 0.0146 }_{ -0.0146 }$                                & $0.3570 ^{+0.0082}_{ -0.0099}$                                  & $ 0.4632 ^{+ 0.0369 }_{ -0.0413 }$                                            & $ 0.1961 ^{+ 0.0042 }_{ -0.0039 }$                                          & $ 0.3786 ^{+ 0.0389 }_{ -0.0366 }$                                 \\ \vspace{0.1cm} 
$q_{1,2}$                                                         & $ 0.0335 ^{+ 0.0407 }_{ -0.0242 }$                                            & $ 0.1113 ^{+ 0.0121 }_{ -0.0112 }$                                  & $ 0.3112 ^{+ 0.0179 }_{ -0.0169 }$                                & $0.2961^{ +0.0085}_{ -0.0068}$                                  & $ 0.3046 ^{+ 0.0518 }_{ -0.0460 }$                                            & $ 0.9959 ^{+ 0.0031 }_{ -0.0068 }$                                          & $ 0.7549 ^{+ 0.0938 }_{ -0.1037 }$                                 \\ \vspace{0.1cm} 
$q_{2,1}$                                                         & $ 0.4027 ^{+ 0.0249 }_{ -0.0710 }$                                            & $ 0.3662 ^{+ 0.0269 }_{ -0.0259 }$                                  & $ 0.3991 ^{+ 0.0265 }_{ -0.0260 }$                                & $0.3436^{+0.1530}_{-0.1270}$                                    & $ 0.5470 ^{+ 0.0680 }_{ -0.0509 }$                                            & $ 0.0424 ^{+ 0.0076 }_{ -0.0070 }$                                          & $ 4056 ^{+ 0.0711 }_{ -0.0553 }$                                 \\ \vspace{0.1cm} 
$q_{2,2}$                                                         & $ 0.0905 ^{+ 0.1343 }_{ -0.0579 }$                                            & $ 0.2377 ^{+ 0.0226 }_{ -0.0213 }$                                  & $ 0.3220 ^{+ 0.0329 }_{ -0.0306 }$                                & $0.3189^{ +0.3426}_{ -0.2208}$                                  & $ 0.2275 ^{+ 0.0427 }_{ -0.0381 }$                                            & $ 0.9380 ^{+ 0.0459 }_{ -0.0923 }$                                          & $ 0.2689 ^{+ 0.2181}_{ -0.1750 }$                                 \\ \vspace{0.1cm} 
$M_1 + M_2~(M_{\odot})$                                           & $ 2.903 ^{+ 0.068 }_{ -0.064 }$                                               & $ 2.765 ^{+ 0.052 }_{ -0.052 }$                                     & $ 2.604 ^{+ 0.038 }_{ -0.038 }$                                   & \nodata                                                         & $ 2.236 ^{+ 0.025 }_{ -0.026 }$                                               & $ 1.645 ^{+ 0.017 }_{ -0.017 }$                                             & $ 2.040 ^{+ 0.073 }_{ -0.074 }$                                    \\ \vspace{0.1cm} 
$M_2/M_1$                                                         & $ 0.863 ^{+ 0.014 }_{ -0.013 }$                                               & $ 0.960 ^{+ 0.015 }_{ -0.015 }$                                     & $ 0.723 ^{+ 0.009 }_{ -0.009 }$                                   & \nodata                                                         & $ 0.969 ^{+ 0.007 }_{ -0.008 }$                                               & $ 0.746 ^{+ 0.005 }_{ -0.005 }$                                             & $ 1.030 ^{+ 0.026 }_{ -0.025 }$                                    \\ \vspace{0.1cm} 
$R_1 + R_2~(R_{\odot})$                                           & $ 3.679 ^{+ 0.031 }_{ -0.033 }$                                               & $ 3.110 ^{+ 0.019 }_{ -0.020 }$                                     & $ 2.746 ^{+ 0.013 }_{ -0.013 }$                                   & $2.8231^{ +0.0010}_{ -0.0009}$                                  & $ 2.060 ^{+ 0.008 }_{ -0.008 }$                                               & $ 1.525 ^{+ 0.005 }_{ -0.005 }$                                             & $ 2.632 ^{+ 0.034 }_{ -0.033 }$                                    \\ \vspace{0.1cm} 
$R_2/R_1$                                                         & $ 0.839 ^{+ 0.055 }_{ -0.067 }$                                               & $ 0.879 ^{+ 0.008 }_{ -0.007 }$                                     & $ 0.5708 ^{+ 0.0003 }_{ -0.0003 }$                                & $0.3283 ^{+0.0002}_{-0.0002}$                                   & $ 0.993 ^{+ 0.020 }_{ -0.018 }$                                               & $ 0.679 ^{+ 0.003 }_{ -0.003 }$                                             & $ 1.222 ^{+ 0.113 }_{ -0.153 }$                                    \\ \vspace{0.1cm} 
$b$\tablenotemark{d}                                              & $ 1.5146 ^{+ 0.0473 }_{ -0.0576 }$                                            & $ 1.2407 ^{+ 0.0081 }_{ -0.0075 }$                                  & $ 0.3860 ^{+ 0.0005 }_{ -0.0005 }$                                & $0.3413^{ +0.0009}_{ -0.0008}$                                  & $ 0.7215 ^{+ 0.0079 }_{ -0.0076 }$                                            & $ 0.5533 ^{+ 0.0041 }_{ -0.0042 }$                                          & $ 1.9034 ^{+ 0.0960 }_{ -0.1341 }$                                 \\ \vspace{0.1cm} 
$e$                                                               & $ 6.7\mathrm{e}\minus5 ^{+ 9.6\mathrm{e}\minus4 }_{ -6.2\mathrm{e}\minus5 }$  & $ 0.63462 ^{+ 0.00001 }_{ -0.00001 }$                               & $ 0.2392 ^{+ 0.0004 }_{ -0.0004 }$                                & $0.00064 ^{+0.00045}_{-0.00033}$                                & $ 3.4\mathrm{e}\minus4 ^{+ 1.3\mathrm{e}\minus3 }_{ -2.6\mathrm{e}\minus4 }$  & $ 0.0127 ^{+ 0.0005 }_{ -0.0004 }$                                          & $ 0.2510 ^{+ 0.0005 }_{ -0.0002 }$                                 \\ \vspace{0.1cm} 
$\ln{\left(\sigma_{\textrm{LC, sys}}\right)}$                     & $ -8.76 ^{+ 0.01 }_{ -0.01 }$                                                 & $ -8.84 ^{+ 0.01 }_{ -0.01 }$                                       & $ -8.48 ^{+ 0.01 }_{ -0.01 }$                                     & $-8.419   ^{ +0.006 }_{ -0.006 }$                               & $ -6.81 ^{+ 0.01 }_{ -0.01 }$                                                 & $ -7.54 ^{+ 0.01 }_{ -0.01 }$                                               & $ -8.12 ^{+ 0.01 }_{ -0.01 }$                                      \\ \vspace{0.1cm} 
$k_0~(\textrm{m s}^{-1})$                                         & $ 61668.8 ^{+ 532.1 }_{ -488.9 }$                                             & $ 93945.5 ^{+ 139.6 }_{ -142.6 }$                                   & $ 72332.2 ^{+ 213.8 }_{ -212.4 }$                                 & $ 68694.7 ^{ +144.0 }_{ -147.0 }$                               & $ 91275.4 ^{+ 248.5 }_{ -256.1 }$                                             & $ 89867.7 ^{+ 127.1 }_{ -125.2 }$                                           & $ 44547.4 ^{+ 633.7 }_{ -627.5 }$                                  \\ \vspace{0.1cm} 
$\ln{\left(\sigma_{\textrm{RV, sys}}~(\textrm{m s}^{-1})\right)}$ & $ 7.40 ^{+ 0.26 }_{ -0.23 }$                                                  & $ 6.54 ^{+ 0.14 }_{ -0.13 }$                                        & $ 7.04 ^{+ 0.12 }_{ -0.11 }$                                      & $ 6.27    ^{ +0.23  }_{ -0.20  }$                               & $ 6.65 ^{+ 0.29 }_{ -0.23 }$                                                  & $ 6.82 ^{+ 0.10 }_{ -0.10 }$                                                & $ 8.28 ^{+ 0.12 }_{ -0.11 }$                                       
\enddata
\tablenotetext{a}{Broad Gaussian prior on flux ratio used based on BF results ($\sigma=20$\% BF value).}
\tablenotetext{b}{$f_M$ is the mass function as defined in \S \ref{keblat}; instead of fitting directly for $M_1$ and $M_2$, we used $f_M$ for KIC 6449358 because it is a single-lined spectroscopic binary.}
\tablenotetext{c}{The fit zeropoint for the time of primary eclipse is in units of BKJD (BJD - 2454833). The primary eclipse is defined here as the deeper of the two. This differs from the KEBC \citep{Kirk_2016} primary eclipse definition for KIC 6864859 only because the two eclipses have very similar depths.}
\tablenotetext{d}{The impact parameter $b$ is defined as $a \cos i/R_1$, where $q^1_1, q^1_2$ and $q^2_1, q^2_2$ are triangularized quadratic limb darkening coefficients for star 1 and star 2 (see Section~\ref{keblat} for details). We use natural log for the systematic LC and RV error terms for fitting flexibility.}
\end{deluxetable*}
\end{rotatetable*}

%% file: table3.tex
\begin{deluxetable*}{lccccccc}
\tablecolumns{8}
\tablewidth{0pt}
\tablecaption{Parameters related to RV extraction and temperature estimates\label{table3}}
\centering
\tablenum{3}
\tablehead{ \colhead{} & 
    \colhead{KIC 5285607} & \colhead{KIC 6864859} &
    \colhead{KIC 6778289} & \colhead{KIC 6449358} & \colhead{KIC 4285087} & \colhead{KIC 6131659} & \colhead{KIC 6781535}
}
\startdata
\vspace{0.1cm}
BF Flux Ratio ($F_2/F_1$) &    $0.620\pm0.027$ & $0.811\pm0.028$ &
                   $0.462\pm0.029$ & $0.392\pm0.014$ & $0.997\pm0.023$ & 
                   $0.648\pm0.028$ & $1.253\pm0.115$ \\ \vspace{0.1cm}
ASPCAP $T_{\textrm{eff}}$ (K)\tablenotemark{a} &    $6495\pm156$ & $6417\pm159$ &
                   $6572\pm162$ & $6237\pm179$ & $5664\pm146$ & 
                   $4845\pm98$ & $5749\pm125$ \\ \vspace{0.1cm}
Gaia parallax (mas)\tablenotemark{b} &    $1.254\pm0.0216$ & $1.4897\pm0.0241$ &
                   $0.9093\pm0.0222$ & $1.1974\pm0.0264$ & $1.619\pm0.0312$ & 
                   $-0.5117\pm1.0713$ & \nodata \\ \vspace{0.1cm}
Gaia distance (pc)\tablenotemark{b} &    $799\pm14$ & $671\pm11$ &
                   $1100\pm27$ & $835\pm18$ & $617\pm12$ & 
                   \nodata & \nodata \\ \vspace{0.1cm}
$\log g_1$ (cgs)\tablenotemark{c} &    $4.028\pm0.013$ & $4.150\pm0.010$ &
                   $4.133\pm0.008$ & \nodata & $4.454\pm0.006$ & 
                   $4.496\pm0.005$ & $4.161\pm0.019$ \\ \vspace{0.1cm}
$\log g_2$ (cgs)\tablenotemark{c} &    $4.118\pm0.015$ & $4.244\pm0.012$ &
                   $4.479\pm0.010$ & \nodata & $4.478\pm0.007$ & 
                   $4.705\pm0.006$ & $4.253\pm0.021$ \\ \vspace{0.1cm}
$T_{\textrm{eff}}$ offset (K)\tablenotemark{d} & $350$ & $80$ &
                   $250$ & \nodata & $20$ & 
                   $350$ & $100$ \\ \vspace{0.1cm}
Adopted $T_{\textrm{eff}, 1}$ (K) &    $6845\pm328$ & $6497\pm159$ &
                   $6822\pm162$ & $6737\pm178$ & $5689\pm146$ & 
                   $5195\pm98$ & $5849\pm125$ \\ \vspace{0.1cm}
Adopted $T_{\textrm{eff}, 2}$ (K) &    $6716\pm293$ & $6541\pm283$ &
                   $7265\pm440$ & $8788\pm658$ & $5735\pm105$ & 
                   \nodata & \nodata
\enddata
\tablenotetext{a}{DR14 \citep{Garcia-Perez_2016}}
\tablenotetext{b}{\citet{Bailer-Jones_2018}}
\tablenotetext{c}{Computed directly from $M$ and $R$ as reported in Table \ref{table2}.}
\tablenotetext{d}{\citet{ElBadry_2017}}
\end{deluxetable*}

%% file: appendixA.tex
\section{Stellar Radial Velocities}
\label{appendixA}

Here, we include one Broadening Function plot for each of the seen targets. These illustrate how we measured radial velocities for each component of the spectroscopic binaries from APOGEE visit spectra, as discussed in Section~\ref{apogee}. Note each figure uses uncorrected RV on the abscissa, before barycentric corrections have been applied. The final corrected RVs with uncertainties are reported in Table~\ref{RVtable} below.

\begin{figure}[ht!]
\centering
\includegraphics[width=0.9\columnwidth]{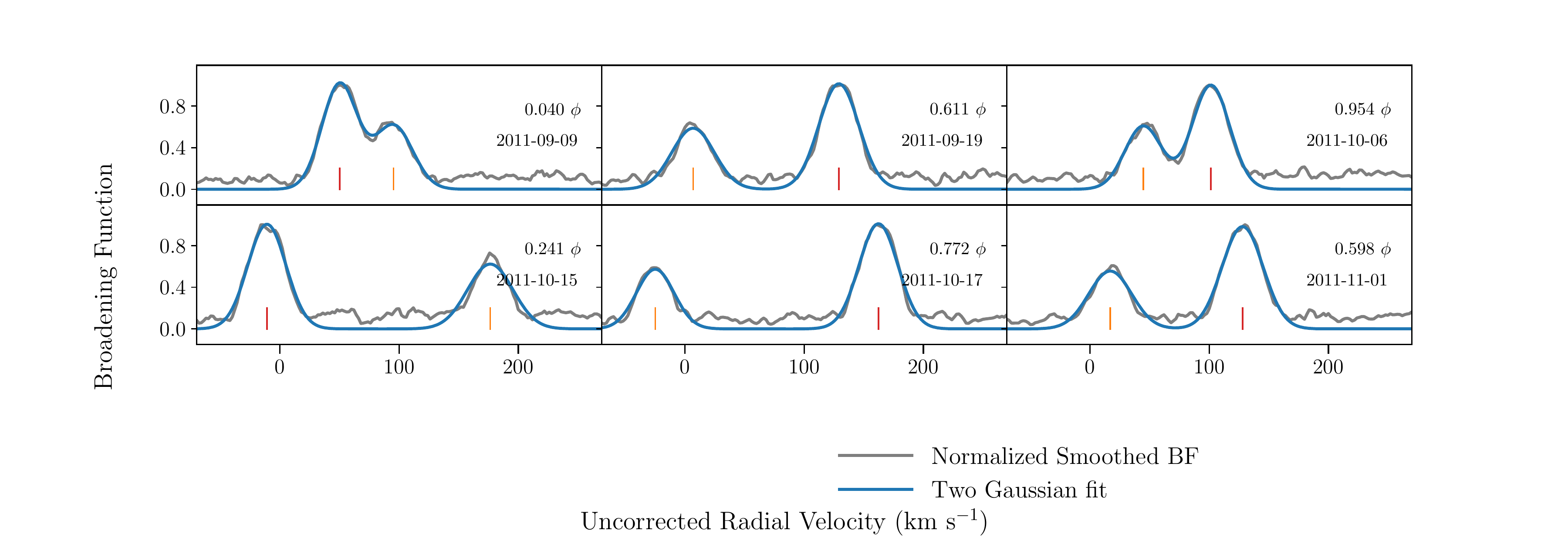}
\caption{{BF plots for KIC 5285607. 
The normalized smoothed BF is shown in grey while the Gaussian fits are
modeled in blue. Uncorrected radial velocities are shown on the abscissa
in km~s$^{-1}$, and arbitrary amplitude of the BF on the vertical axis. 
In this case the
primary (red) is distinguishable from the secondary (orange), and the
visit spectra were well separated over the coarse of the observations.
{\label{fig_5285607bf}}
}}
\end{figure}

\begin{figure}[ht!]
\centering
\includegraphics[width=0.9\columnwidth]{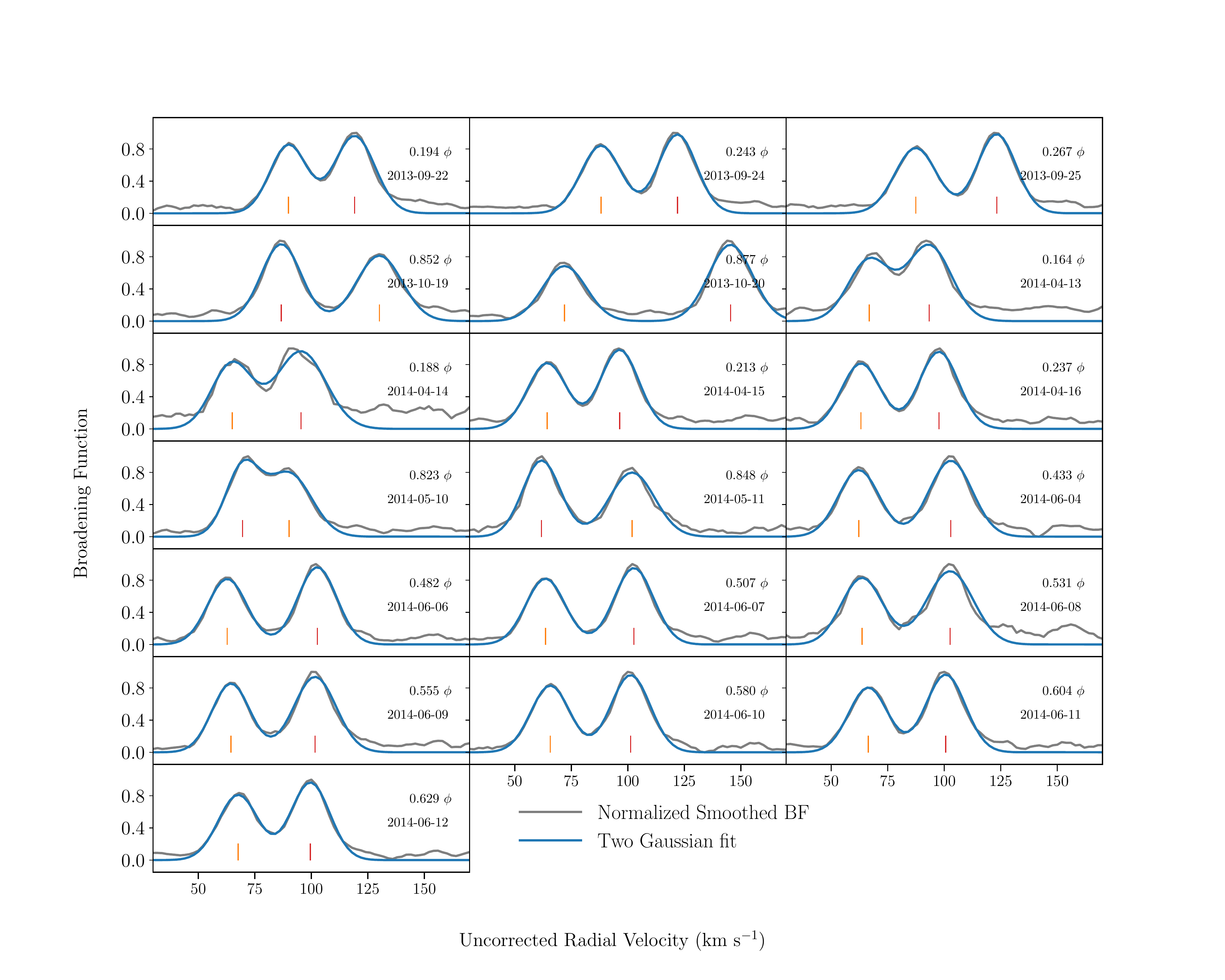}
\caption{{Same as Figure~\ref{fig_5285607bf} but for KIC 6864859. In this case the
primary (red) and secondary (orange) are less distinguishable due to the
greater radius of the primary. Observations often were within a day of
one another in their respective visit sets. Some visits were removed due
to the presence of noise that could not be eliminated with our despiking
method. 
{\label{fig_6864859bf}}
}}
\end{figure}

\begin{figure}[ht!]
\centering
\includegraphics[width=\columnwidth]{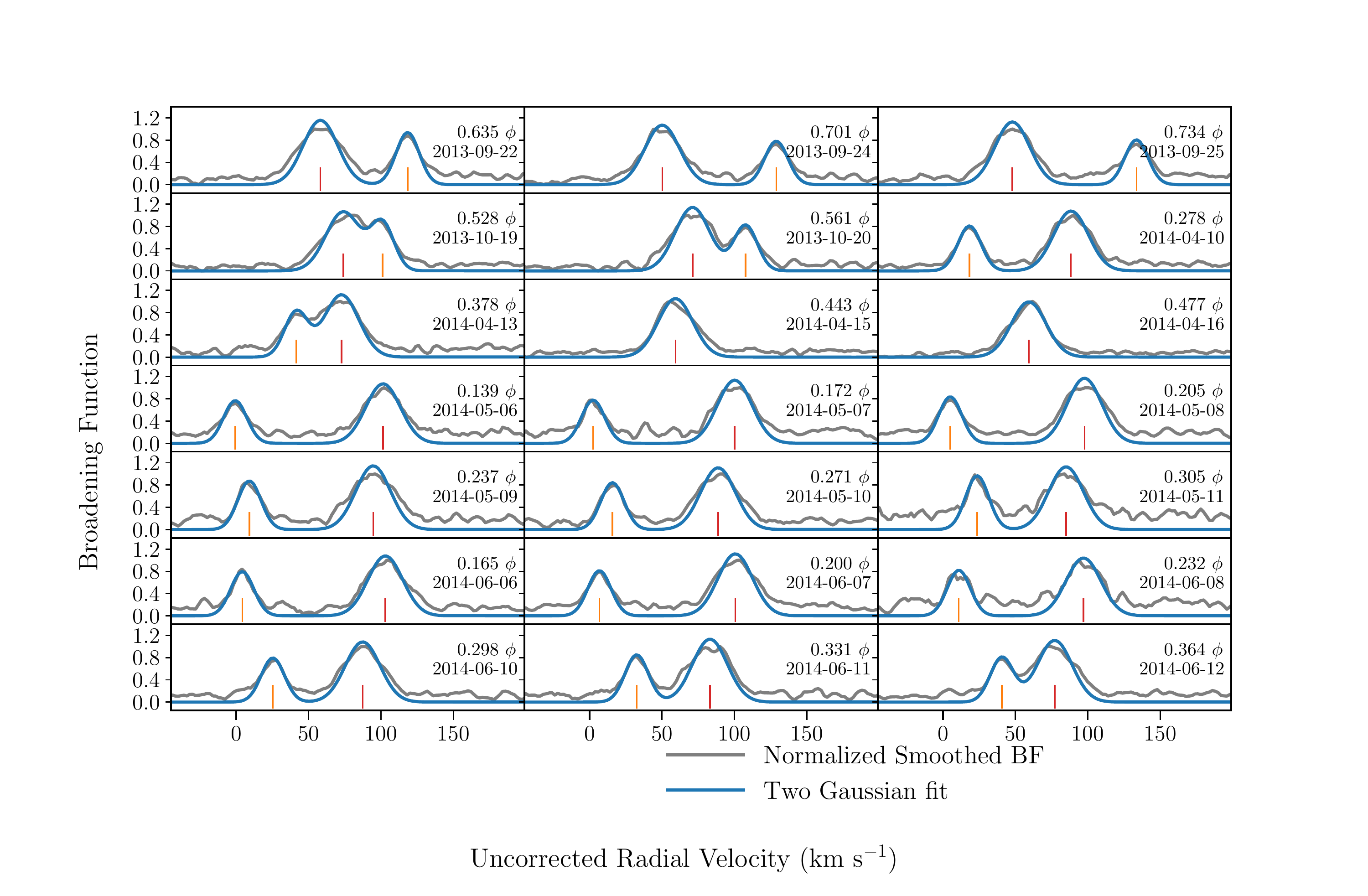}
\caption{{Same as Figure~\ref{fig_5285607bf} but for KIC 6778289. In this case the
primary (red) is slightly distinguishable from the secondary (orange),
and the visit spectra were well separated over the coarse of the
observations. Some visits were removed due to excess noise after having
been ran through our despiking method. 
{\label{fig_6778289bf}}
}}
\end{figure}

\begin{figure}[ht!]
\centering
\includegraphics[width=\columnwidth]{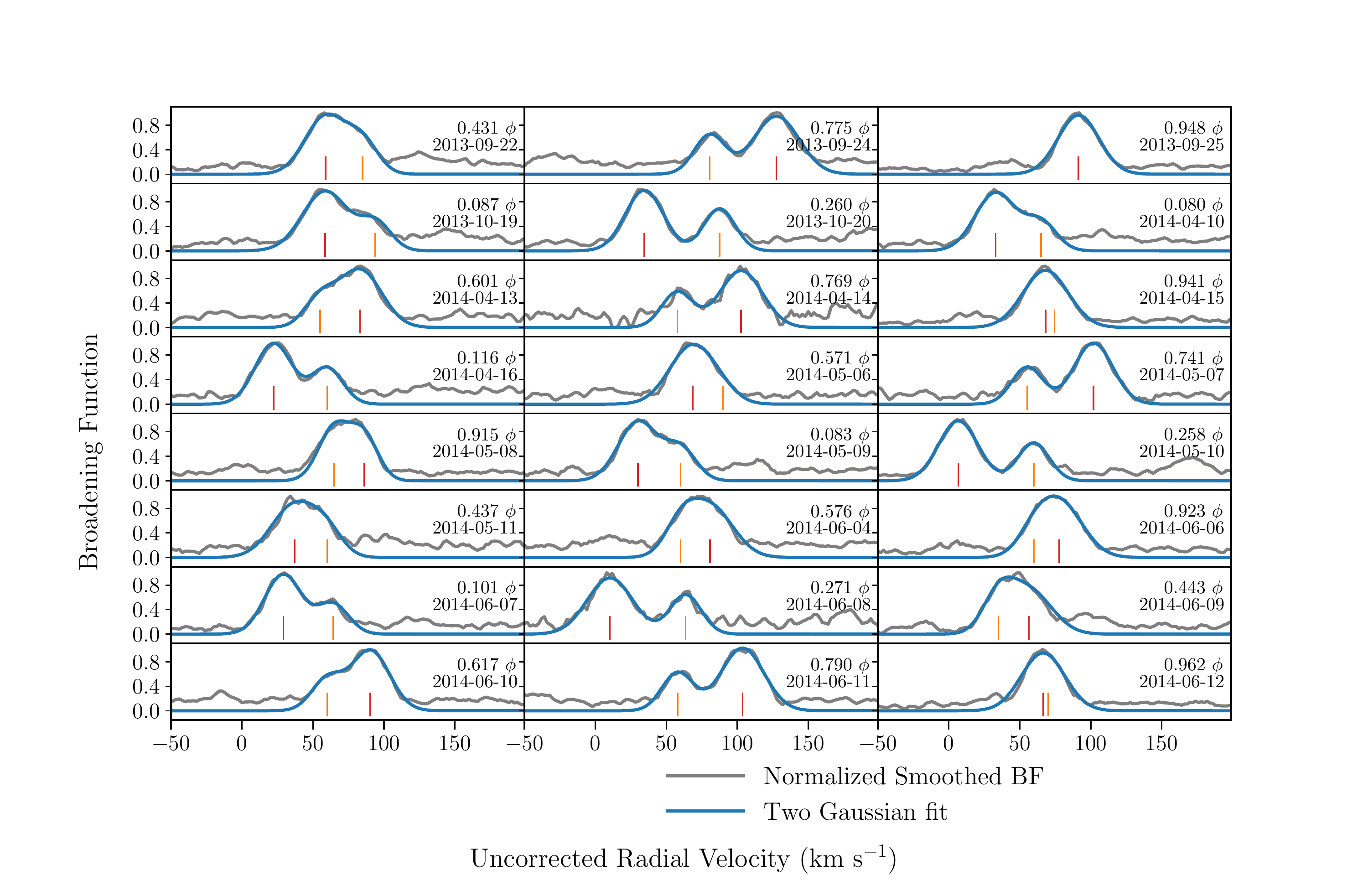}
\caption{{Same as Figure~\ref{fig_5285607bf} but for KIC 6449358. In this case the
primary (red) is easily distinguishable from the secondary (orange).
This target had visits in which the primary and secondary were very
close together but not directly on top of one another, in these
occurrences error is more pronounced in the radial velocities extracted.
{\label{fig_6449358bf}}
}}
\end{figure}

\begin{figure}[ht!]
\centering
\includegraphics[width=\columnwidth]{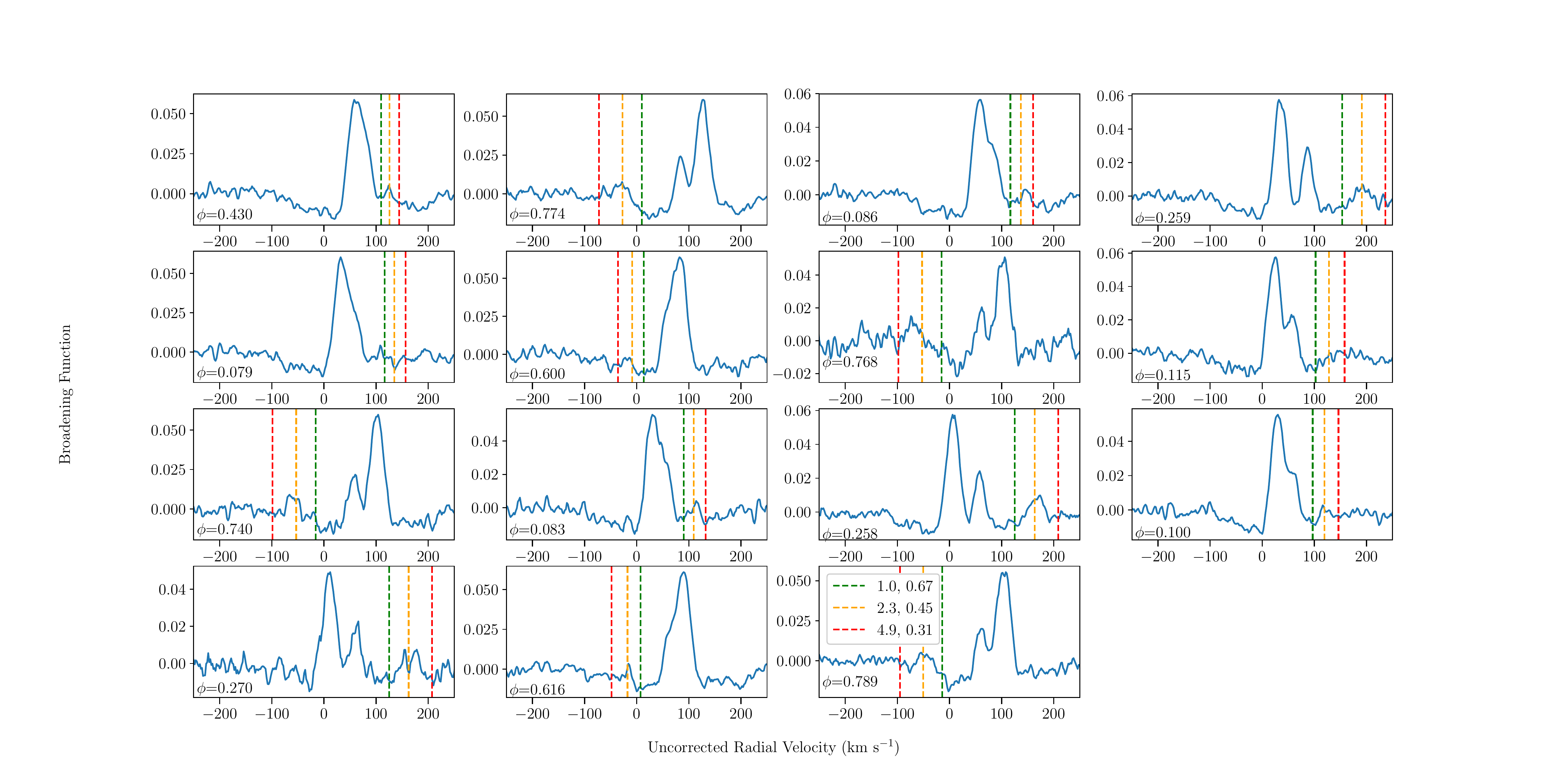}
\caption{{BF (blue solid line; zoomed out in each panel to $\pm$250 km~s$^{-1}$) of KIC 6449358 overplotted with expected locations of the secondary component's BF peak corresponding to various $M_1+M_2, M_2/M_1$ combinations (green, orange, red dashed lines). In a few of the visits, tentative BF peaks coincide with the secondary's expected locations for $M_1+M_2 \sim 2.3, M_2/M_1 \sim 0.45$ (see, e.g., panels corresponding to phase=0.741, 0.083, 0.258, 0.259, 0.430). In general, however, additional BF ``peaks" are lost in the noise of the spectra. 
{\label{fig_6449358bf_forwardmodel}}
}}
\end{figure}

\begin{figure}[ht!]
\centering
\includegraphics[width=\columnwidth]{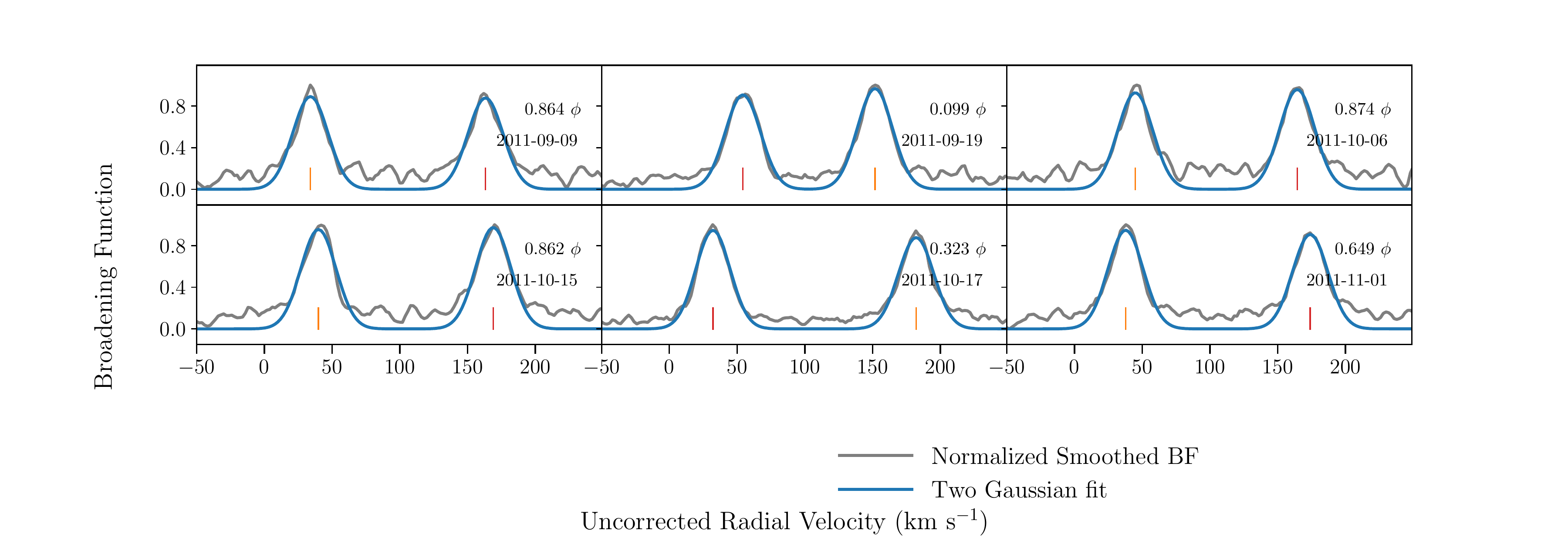}
\caption{{Same as Figure~\ref{fig_5285607bf} but for KIC 4285087. In this case the
primary (red) is easily distinguishable from the secondary (orange). 
{\label{fig_4285087bf}}
}}
\end{figure}

\begin{figure}[ht!]
\centering
\includegraphics[width=\columnwidth]{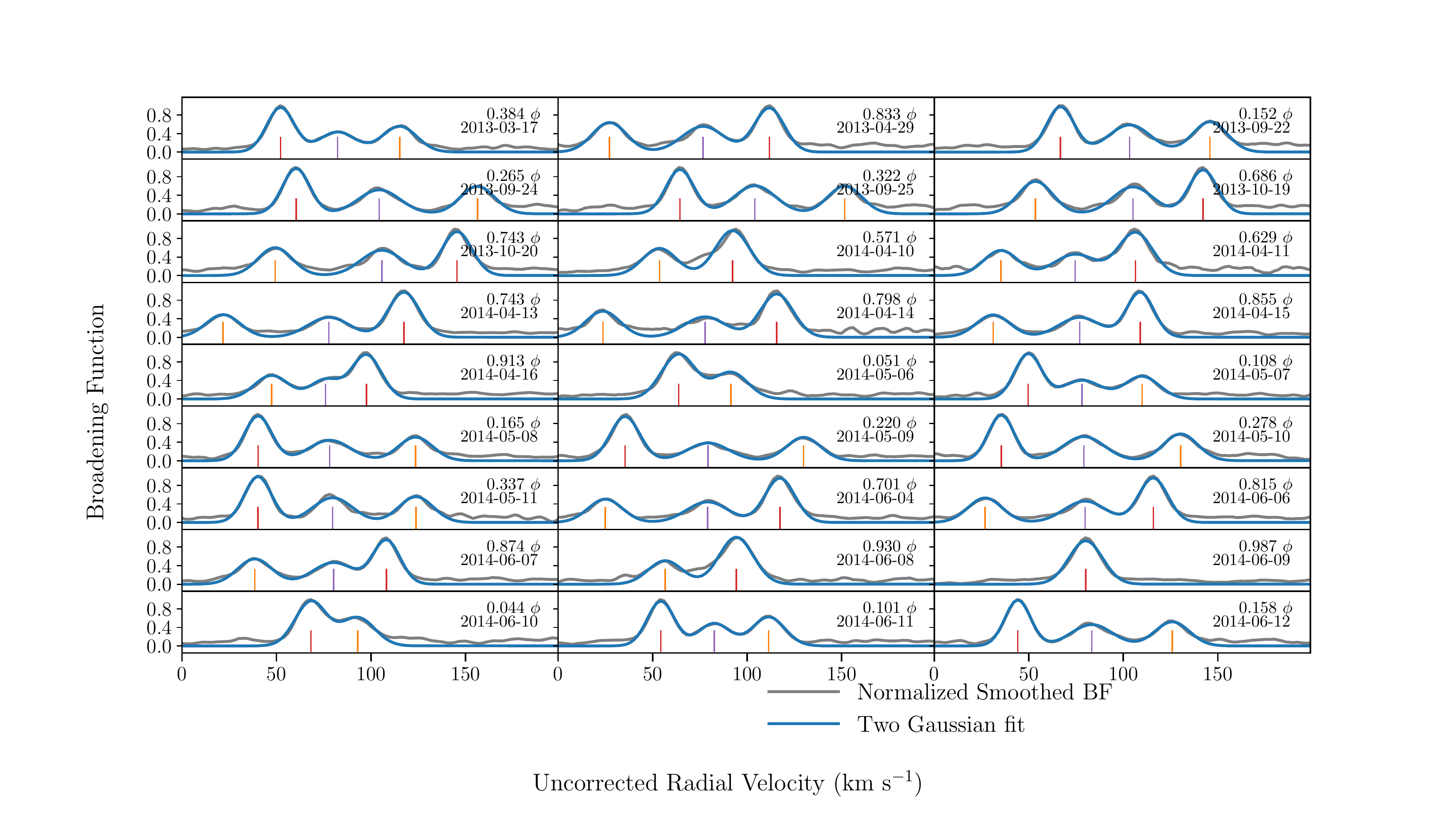}
\caption{{Same as Figure~\ref{fig_5285607bf} but for KIC 6131659. In this case the
primary (red) is very distinguishable from the secondary (orange). A
tertiary (purple) member is visible but it does not show variance in its
radial velocity component. In some panels the tertiary
is not visible because it is very near to or within the primary or
secondary peak. 
{\label{fig_6131659bf}}
}}
\end{figure}

\begin{figure}[ht!]
\centering
\includegraphics[width=\columnwidth]{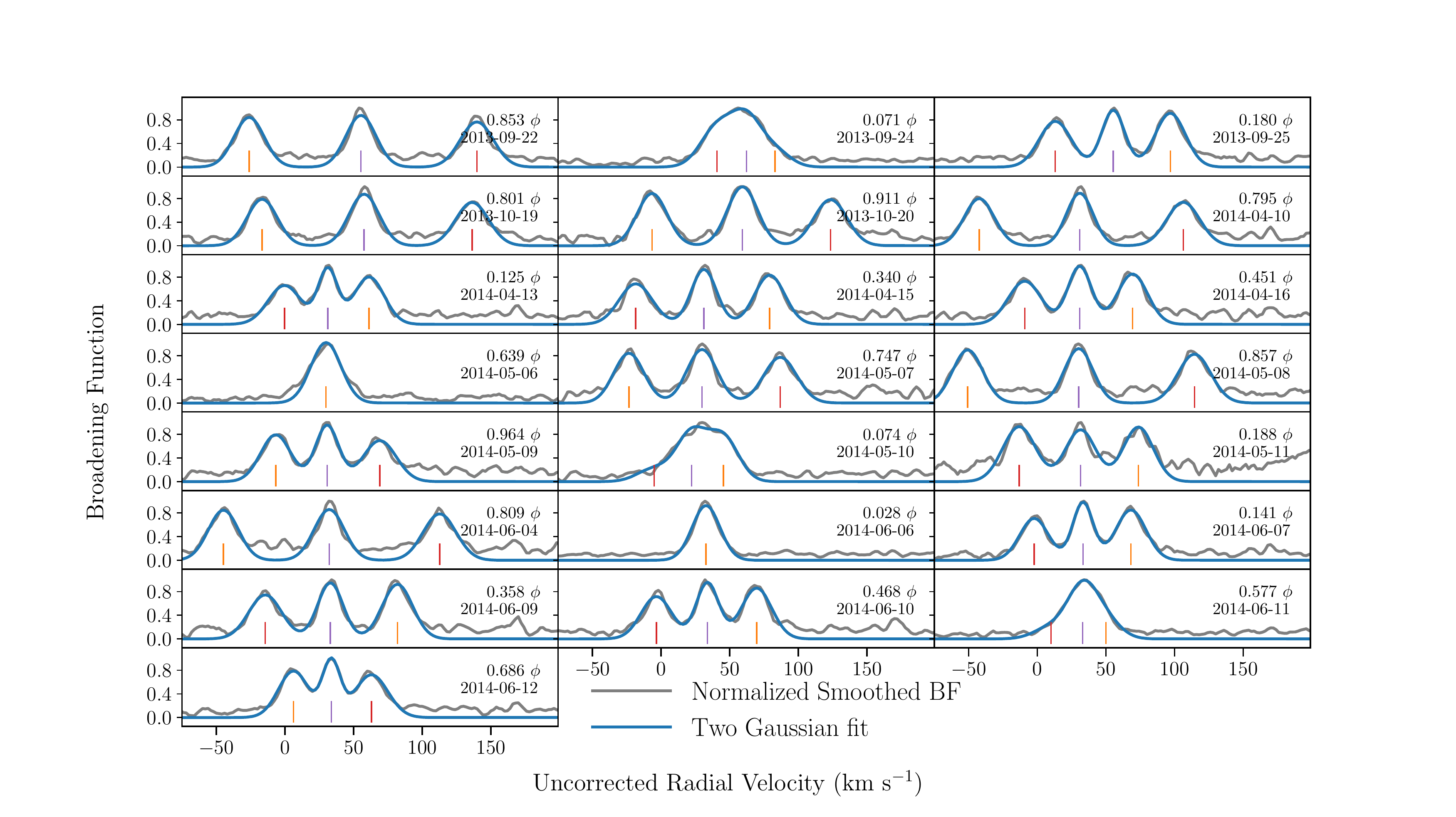}
\caption{{Same as Figure~\ref{fig_5285607bf} but for KIC 6781535. In most panels, the primary (red) is very distinguishable from the secondary (orange). A
tertiary member is present (purple) but is not RV
variant. 
{\label{fig_6781535bf}}
}}
\end{figure}

\clearpage

\input{table_rvs.tex}

%% file: table_rvs.tex
\startlongtable
\begin{deluxetable}{cccc}
\label{RVtable}
\tablecolumns{4}
\tablewidth{0pt}
\tablecaption{Measured radial velocities from APOGEE spectra}
\centering
\tablenum{4}
\tablehead{
\colhead{Time (BJD)} & \colhead{Orbital Phase} & \colhead{$v_1$ ($\rm{km \ s}^{-1}$)} & \colhead{$v_2$ ($\rm{km \ s}^{-1}$)}
}
\startdata
\cutinhead{KIC 5285607}
2455813.69967  & 0.040 & $ 41.404 \pm  0.101$ & $ 86.618  \pm  0.173$ \\
2455823.72647  & 0.611 & $118.406 \pm  0.061$ & $ -3.961  \pm  0.108$ \\
2455840.66112  & 0.954 & $ 88.218 \pm  0.067$ & $ 31.481  \pm  0.109$ \\
2455849.57835  & 0.241 & $-24.927 \pm  0.060$ & $162.515  \pm  0.105$ \\
2455851.64874  & 0.772 & $148.087 \pm  0.060$ & \nodata             \\
2455866.56945  & 0.598 & $113.109 \pm  0.063$ & $  2.025  \pm  0.114$ \\
\cutinhead{KIC 6864859}
2456557.73263 & 0.194 & $107.492 \pm  0.058$ &  \nodata \\
2456559.72256 & 0.243 & $110.939 \pm  0.079$ &  $ 78.670 \pm  0.069$ \\
2456560.72029 & 0.267 & $112.129 \pm  0.062$ &  $ 77.683 \pm  0.056$ \\
2456584.63147 & 0.852 & $ 74.820 \pm  0.060$ &  $117.933 \pm  0.063$ \\
2456585.62998 & 0.877 & $ 57.225 \pm  0.064$ & $133.366 \pm  0.052$ \\ 
2456760.90512 & 0.164 & $106.432 \pm  0.072$ &  $ 80.051 \pm  0.077$ \\
2456761.87222 & 0.188 & $108.904 \pm  0.073$ &  $ 78.499 \pm  0.079$ \\
2456762.86801 & 0.213 & $109.800 \pm  0.057$ &  $ 77.234 \pm  0.048$ \\
2456763.88053 & 0.237 & $111.281 \pm  0.053$ &  $ 76.175 \pm  0.054$ \\
2456787.80872 & 0.823 & $ 81.571 \pm  0.078$ &  $104.978 \pm  0.091$ \\
2456788.84246 & 0.848 & $ 76.011 \pm  0.053$ &  $116.930 \pm  0.062$ \\
2456812.75131 & 0.433 & $114.944 \pm  0.053$ &  $ 73.779 \pm  0.051$ \\
2456814.75480 & 0.482 & $114.257 \pm  0.049$ &  $ 73.825 \pm  0.050$ \\
2456815.78485 & 0.507 & $113.825 \pm  0.053$ &  $ 73.899 \pm  0.049$ \\
2456816.76560 & 0.531 & $113.034 \pm  0.061$ &  $ 73.480 \pm  0.061$ \\
2456817.76131 & 0.555 & $112.440 \pm  0.056$ &  $ 74.591 \pm  0.048$ \\
2456818.76390 & 0.580 & $111.523 \pm  0.056$ &  $ 75.498 \pm  0.054$ \\
2456819.76154 & 0.604 & $110.927 \pm  0.058$ &  $ 75.851 \pm  0.053$ \\
2456820.75533 & 0.629 & $109.049 \pm  0.059$ &  $ 75.654 \pm  0.066$ \\
\cutinhead{KIC 6778289}  
2456557.73261 & 0.635 & $ 47.229 \pm  0.045$ &  $107.444  \pm  0.045$ \\
2456559.72254 & 0.701 & $ 38.941 \pm  0.049$ &  $117.686  \pm  0.054$ \\
2456560.72027 & 0.734 & $ 36.535 \pm  0.046$ &  $122.226  \pm  0.053$ \\
2456584.63145 & 0.528 & $ 60.403 \pm  0.059$ &  $ 87.443  \pm  0.058$ \\
2456585.62996 & 0.561 & $ 57.486 \pm  0.047$ &  $ 94.149  \pm  0.053$ \\
2456757.89237 & 0.278 & $101.533 \pm  0.048$ &  $ 31.407  \pm  0.053$ \\
2456760.90514 & 0.378 & $ 85.993 \pm  0.051$ &  $ 54.705  \pm  0.056$ \\
2456762.86803 & 0.443 & $ 72.743 \pm  0.049$ &  \nodata   \\
2456763.88055 & 0.477 & $ 72.718 \pm  0.052$ &  \nodata   \\
2456783.83502 & 0.139 & $115.038 \pm  0.049$ &  $ 12.937  \pm  0.055$ \\
2456784.82136 & 0.172 & $113.771 \pm  0.046$ &  $ 16.022  \pm  0.055$ \\
2456785.82484 & 0.205 & $111.263 \pm  0.044$ &  $ 18.581  \pm  0.051$ \\
2456786.79785 & 0.237 & $108.112 \pm  0.045$ &  $ 22.629  \pm  0.048$ \\
2456787.80874 & 0.271 & $102.199 \pm  0.047$ &  $ 29.240  \pm  0.050$ \\
2456788.84248 & 0.305 & $ 98.364 \pm  0.046$ &  $ 36.995  \pm  0.044$ \\
2456814.75483 & 0.165 & $113.882 \pm  0.048$ &  $ 15.109  \pm  0.053$ \\
2456815.78487 & 0.200 & $111.327 \pm  0.047$ &  $ 17.508  \pm  0.052$ \\
2456816.76563 & 0.232 & $107.622 \pm  0.050$ &  $ 21.517  \pm  0.052$ \\
2456818.76392 & 0.298 & $ 97.639 \pm  0.048$ &  $ 35.617  \pm  0.053$ \\
2456819.76156 & 0.331 & $ 93.260 \pm  0.046$ &  $ 42.669  \pm  0.050$ \\
2456820.75535 & 0.364 & $ 87.120 \pm  0.048$ &  $ 50.657  \pm  0.054$ \\
\cutinhead{KIC 6449358}
2456557.73275 & 0.431 & $ 47.972 \pm  0.092$ &  \nodata \\ 
2456559.72268 & 0.775 & $116.332 \pm  0.066$ &  \nodata \\ 
2456584.63158 & 0.087 & $ 44.828 \pm  0.067$ &  \nodata \\ 
2456585.63008 & 0.260 & $ 20.640 \pm  0.056$ &  \nodata \\ 
2456757.89224 & 0.080 & $ 46.407 \pm  0.069$ &  \nodata \\ 
2456760.90501 & 0.601 & $ 96.766 \pm  0.090$ &  \nodata \\ 
2456761.87212 & 0.769 & $116.275 \pm  0.066$ &  \nodata \\ 
2456763.88043 & 0.116 & $ 36.021 \pm  0.056$ &  \nodata \\ 
2456784.82126 & 0.741 & $115.987 \pm  0.057$ &  \nodata \\ 
2456787.80865 & 0.258 & $ 20.638 \pm  0.057$ &  \nodata \\ 
2456815.78483 & 0.101 & $ 40.477 \pm  0.069$ &  \nodata \\ 
2456816.76558 & 0.271 & $ 21.538 \pm  0.066$ &  \nodata \\ 
2456818.76389 & 0.617 & $101.418 \pm  0.065$ &  \nodata \\ 
2456819.76152 & 0.790 & $114.484 \pm  0.060$ &  \nodata \\ 
\cutinhead{KIC 4285087}  
2455813.69984 & 0.864 & $154.531 \pm  0.062$ & $ 25.294 \pm  0.060$  \\
2455823.72663 & 0.099 & $ 43.398 \pm  0.060$ & $140.891 \pm  0.056$  \\
2455840.66127 & 0.874 & $150.887 \pm  0.057$ & $ 31.327 \pm  0.058$  \\
2455849.57849 & 0.862 & $154.496 \pm  0.056$ & $ 25.174 \pm  0.057$  \\
2455851.64888 & 0.323 & $ 17.397 \pm  0.057$ & $167.332 \pm  0.062$  \\
2455866.56955 & 0.649 & $158.478 \pm  0.060$ & $ 22.212 \pm  0.057$  \\
\cutinhead{KIC 6131659}
2456368.99876 & 0.384 & $ 62.544 \pm  0.048$ & $125.604 \pm  0.091$   \\ 
2456411.91961 & 0.833 & $125.989 \pm  0.045$ & $ 41.382 \pm  0.070$   \\ 
2456557.73279 & 0.152 & $ 55.739 \pm  0.042$ & $134.879 \pm  0.069$   \\ 
2456559.72271 & 0.265 & $ 49.076 \pm  0.041$ & $145.012 \pm  0.076$   \\ 
2456560.72045 & 0.322 & $ 52.988 \pm  0.043$ & $140.191 \pm  0.075$   \\ 
2456584.63160 & 0.686 & $128.199 \pm  0.045$ & $ 39.531 \pm  0.064$   \\ 
2456585.63010 & 0.743 & $131.456 \pm  0.045$ & $ 35.183 \pm  0.075$   \\ 
2456757.89221 & 0.571 & $105.771 \pm  0.047$ & $ 67.116 \pm  0.077$   \\ 
2456758.90157 & 0.629 & $120.004 \pm  0.052$ & $ 48.763 \pm  0.084$   \\ 
2456760.90499 & 0.743 & $131.105 \pm  0.044$ & $ 35.406 \pm  0.092$   \\ 
2456761.87209 & 0.798 & $129.464 \pm  0.048$ & $ 37.476 \pm  0.080$   \\ 
2456762.86788 & 0.855 & $122.805 \pm  0.042$ & $ 45.003 \pm  0.093$   \\ 
2456763.88040 & 0.913 & $111.491 \pm  0.099$ & $ 61.304 \pm  0.105$   \\ 
2456783.83490 & 0.051 & $ 78.027 \pm  0.051$ & $105.677 \pm  0.086$   \\ 
2456784.82125 & 0.108 & $ 63.945 \pm  0.053$ & $124.186 \pm  0.111$   \\ 
2456785.82473 & 0.165 & $ 54.481 \pm  0.042$ & $137.809 \pm  0.089$   \\ 
2456786.79774 & 0.220 & $ 49.620 \pm  0.044$ & $144.068 \pm  0.089$   \\ 
2456787.80864 & 0.278 & $ 49.600 \pm  0.040$ & $144.493 \pm  0.078$   \\ 
2456788.84238 & 0.337 & $ 54.269 \pm  0.041$ & $137.917 \pm  0.082$   \\ 
2456812.75129 & 0.701 & $129.389 \pm  0.044$ & $ 36.993 \pm  0.089$   \\ 
2456814.75479 & 0.815 & $127.626 \pm  0.044$ & $ 38.646 \pm  0.085$   \\ 
2456815.78484 & 0.874 & $119.797 \pm  0.047$ & $ 50.101 \pm  0.083$   \\ 
2456816.76560 & 0.930 & $105.694 \pm  0.045$ & $ 68.143 \pm  0.089$   \\ 
2456817.76131 & 0.987 & $ 91.467 \pm  0.048$ & \nodata                \\ 
2456818.76390 & 0.044 & $ 79.392 \pm  0.046$ & \nodata                \\ 
2456819.76154 & 0.101 & $ 65.338 \pm  0.047$ & $122.387 \pm  0.086$   \\ 
2456820.75533 & 0.158 & $ 55.054 \pm  0.040$ & $136.794 \pm  0.088$   \\ 
\cutinhead{KIC 6781535}  
2456557.73097 & 0.853 &  $128.940 \pm  0.057$ & $-37.069 \pm  0.049$  \\ 
2456559.72097 & 0.071 &  $ 29.463 \pm  0.109$ & $ 69.329 \pm  0.204$  \\ 
2456560.71874 & 0.180 &  $  1.983 \pm  0.057$ & $ 85.464 \pm  0.045$  \\ 
2456584.63091 & 0.801 &  $122.687 \pm  0.060$ & $-30.426 \pm  0.052$  \\ 
2456585.62946 & 0.911 &  $109.784 \pm  0.057$ & $-20.436 \pm  0.047$  \\ 
2456757.89316 & 0.795 &  $119.307 \pm  0.060$ & $-29.253 \pm  0.052$  \\ 
2456760.90580 & 0.125 &  $ 12.366 \pm  0.076$ & $ 75.156 \pm  0.060$  \\ 
2456762.86860 & 0.340 &  $ -4.835 \pm  0.064$ & $ 92.365 \pm  0.049$  \\ 
2456763.88108 & 0.451 &  $  4.541 \pm  0.060$ & $ 82.610 \pm  0.049$  \\ 
2456783.83465 & 0.639 &  \nodata              & $ 43.811 \pm  0.041$   \\ 
2456784.82095 & 0.747 &  $100.401 \pm  0.057$ & $ -9.574 \pm  0.049$  \\ 
2456785.82438 & 0.857 &  $128.143 \pm  0.053$ & $-37.060 \pm  0.046$  \\ 
2456786.79735 & 0.964 &  $ 82.332 \pm  0.064$ & $  7.094 \pm  0.053$  \\ 
2456787.80820 & 0.074 &  $ 23.165 \pm  0.158$ & $ 62.731 \pm  0.119$  \\ 
2456788.84189 & 0.188 &  $  1.170 \pm  0.047$ & $ 86.509 \pm  0.045$  \\ 
2456812.74399 & 0.809 &  $124.321 \pm  0.056$ & $-33.597 \pm  0.049$  \\ 
2456814.75319 & 0.028 &  \nodata              & $ 43.896 \pm  0.046$   \\ 
2456815.78320 & 0.141 &  $  8.643 \pm  0.068$ & $ 79.192 \pm  0.050$  \\ 
2456817.75959 & 0.358 &  $ -3.732 \pm  0.059$ & $ 92.410 \pm  0.045$  \\ 
2456818.76215 & 0.468 &  $  7.142 \pm  0.064$ & $ 80.126 \pm  0.049$  \\ 
2456819.75975 & 0.577 &  $ 20.702 \pm  0.354$ & $ 57.626 \pm  0.204$  \\ 
2456820.75351 & 0.686 &  $ 73.353 \pm  0.070$ & $ 15.883 \pm  0.061$  \\ 
\enddata
\end{deluxetable}